\documentclass[letterpaper,twocolumn,10pt]{article}
\usepackage{usenix}
\usepackage{tikz}
\usepackage{filecontents}
\usepackage{xspace} 
\usepackage{cite}
\usepackage{amsmath,amssymb,amsfonts}
\usepackage{graphicx}
\usepackage{textcomp}
\usepackage{xcolor}
\usepackage{todonotes}
\usepackage{mathtools}
\usepackage{algorithm}
\usepackage[noend]{algpseudocode}
\usepackage{subfig}
\usepackage{tabularx}
\usepackage{booktabs}
\usepackage{makecell}
\usepackage{multirow}
\usepackage[available]{usenixbadges}

\algnewcommand\algorithmicforeach{\textbf{for each}}
\algdef{S}[FOR]{ForEach}[1]{\algorithmicforeach\ #1\ \algorithmicdo}

\usepackage{cleveref}
\usepackage{xcolor,pifont}
	
\usepackage{color, colortbl}

\newcommand{\ournameNoSpace}{GateBreaker} 
\newcommand{\ourname}{\ournameNoSpace\xspace}

\newcommand{\nadd}[1]{{\color{black}{#1}}}

\newcommand*\colourcheck[1]{%
  \expandafter\newcommand\csname #1check\endcsname{\textcolor{#1}{\ding{52}}}%
}

\newcommand*\colourcross[1]{%
  \expandafter\newcommand\csname #1cross\endcsname{\textcolor{#1}{\ding{55}}}%
}

\definecolor{LightCyan}{rgb}{0.88,1,1}
	
\definecolor{Gray}{gray}{0.8}
\colourcheck{green}
\colourcross{red}

\definecolor{darkgreen}{rgb}{0.0, 0.5, 0.0}

\usepackage{listings}
\definecolor{codegreen}{rgb}{0,0.6,0}
\definecolor{codegray}{rgb}{0.5,0.5,0.5}
\definecolor{codepurple}{rgb}{0.58,0,0.82}
\definecolor{backcolour}{rgb}{0.95,0.95,0.92}

\let\othelstnumber=\thelstnumber
\def\createlinenumber#1#2{
    \edef\thelstnumber{%
        \unexpanded{%
            \ifnum#1=\value{lstnumber}\relax
              #2%
            \else}%
        \expandafter\unexpanded\expandafter{\thelstnumber\othelstnumber\fi}%
    }
    \ifx\othelstnumber=\relax\else
      \let\othelstnumber\relax
    \fi
}

\lstdefinestyle{customc}{
  belowcaptionskip=1\baselineskip,
  breaklines=true,
  frame=single,
  xleftmargin=0.35cm,
  xrightmargin=0.15cm,
  numbers=left,
  numbersep=5pt,  
  language=C,
  showstringspaces=false,
  basicstyle=\footnotesize\ttfamily,
  keywordstyle=\bfseries\color{green!40!black},
  commentstyle=\itshape\color{purple!40!black},
  identifierstyle=\color{blue},
  stringstyle=\color{orange},
}

\lstdefinestyle{customcArianeExploit1}{
  breaklines=true,
  frame=single,
  xleftmargin=0.4cm,
  xrightmargin=0.2cm,
  numbers=left,
  numbersep=5pt,  
  language=C,
  showstringspaces=false,
  basicstyle=\footnotesize\ttfamily,
  keywordstyle=\bfseries\color{green!40!black},
  commentstyle=\itshape\color{purple!60!black},
  identifierstyle=\color{blue},
  stringstyle=\color{yellow!50!black},
  morekeywords={asm},
  keywordstyle=[2]\bfseries\color{brown!60!black},
}

\lstdefinestyle{customcArianeExploit}{
  breaklines=true,
  frame=single,
  xleftmargin=0.4cm,
  xrightmargin=0.2cm,
  numbers=left,
  numbersep=5pt,  
  language=C,
  showstringspaces=false,
  basicstyle=\footnotesize\ttfamily,
  keywordstyle=\bfseries\color{blue},
  commentstyle=\itshape\color{green!50!black},
  identifierstyle=\color{black},
  stringstyle=\color{brown},
  morekeywords={asm},
  keywordstyle=[2]\bfseries\color{black},
}

\lstdefinestyle{customlog}{
  breaklines=true,
  frame=single,
  xleftmargin=0.35cm,
  xrightmargin=0.15cm,
  numbers=left,
  numbersep=5pt,  
  language=C,
  showstringspaces=false,
  basicstyle=\footnotesize\ttfamily,
  keywordstyle=\color{blue},
  commentstyle=\itshape\color{purple!40!black},
  identifierstyle=\color{blue},
  stringstyle=\color{orange},
  keywords=[2]{INFO},
  keywords=[3]{ERROR},x
  keywordstyle=[2]\bfseries\color{green!40!black},
  keywordstyle=[3]\bfseries\color{red!500!black},
}

\definecolor{verilogcommentcolor}{RGB}{104,180,104}
\definecolor{verilogkeywordcolor}{RGB}{49,49,255}
\definecolor{verilogsystemcolor}{RGB}{128,0,255}
\definecolor{verilognumbercolor}{RGB}{255,143,102}
\definecolor{verilogstringcolor}{RGB}{160,160,160}
\definecolor{verilogdefinecolor}{RGB}{128,64,0}
\definecolor{verilogoperatorcolor}{RGB}{0,0,128}
\definecolor{pointcolor}{RGB}{192,0,0} 
\lstdefinestyle{prettyverilog}{
   language           = Verilog,
   commentstyle       = \color{verilogcommentcolor},
   alsoletter         = \$'0123456789\`,
   literate           = *{+}{{\verilogColorOperator{+}}}{1}%
                         {-}{{\verilogColorOperator{-}}}{1}%
                         {@}{{\verilogColorOperator{@}}}{1}%
                         {;}{{\verilogColorOperator{;}}}{1}%
                         {*}{{\verilogColorOperator{*}}}{1}%
                         {?}{{\verilogColorOperator{? }}}{1}%
                         {:}{{\verilogColorOperator{:}}}{1}%
                         {<}{{\verilogColorOperator{<}}}{1}%
                         {>}{{\verilogColorOperator{> }}}{1}%
                         {!}{{\verilogColorOperator{!}}}{1}%
                         {^}{{\verilogColorOperator{^}}}{1}%
                         {|}{{\verilogColorOperator{|}}}{1}%
                         {||}{{\verilogColorOperator{|| }}}{1}%
                         {=}{{\verilogColorOperator{= }}}{1}%
                         {==}{{\verilogColorOperator{== }}}{1}%
                         {=>}{{\verilogColorOperator{=> }}}{1}%
                         {[}{{\verilogColorOperator{[}}}{1}%
                         {]}{{\verilogColorOperator{]}}}{1}%
                         {(}{{\verilogColorOperator{(}}}{1}%
                         {)}{{\verilogColorOperator{)}}}{1}%
                         {,}{{\verilogColorOperator{,}}}{1}%
                         {.}{{\verilogColorOperator{.}}}{1}%
                         {~}{{\verilogColorOperator{$\sim$}}}{1}%
                         {\%}{{\verilogColorOperator{\%}}}{1}%
                         {\&}{{\verilogColorOperator{\& }}}{1}%
                         {\&\&}{{\verilogColorOperator{\&\& }}}{1}%
                         {\#}{{\verilogColorOperator{\#}}}{1}%
                         {\ /\ }{{\verilogColorOperator{\ /\ }}}{3}%
                         {\ _}{\ \_}{2}%
                        ,
   morestring         = [s][\color{verilogstringcolor}]{"}{"},%
   identifierstyle    = \color{black},
   vlogdefinestyle    = \color{verilogdefinecolor},
   vlogconstantstyle  = \color{verilognumbercolor},
   vlogsystemstyle    = \color{verilogsystemcolor},
   basicstyle         = \scriptsize\fontencoding{T1}\ttfamily,
  columns=fullflexible, 
   keywordstyle       = \bfseries\color{verilogkeywordcolor},
   morekeywords      = {val, when, port, coverage, unique},
   numbers            = left,
   numbersep          = 5pt,
   tabsize            = 2,
   escapeinside       = {/*!}{!*/},
   upquote            = true,
   sensitive          = true,
   showstringspaces   = false, 
   frame              = single,
   breaklines         = true,
   abovecaptionskip   = 0pt,
   belowcaptionskip   = 2pt,   
   xleftmargin        =0.35cm,
   xrightmargin       =0.15cm,
   captionpos         = t,
   emph               = {Point, Point0, Point1, Point2, Point3, Point4, Point5, Point6, Point7, Point8, Point9},
   emphstyle          =\color{pointcolor},
   emph               = {[2] STVEC,SCOUNTEREN,MSTATUS,MTVEC,ML1_ICACHE_MISS,ML1_DCACHE_MISS,MITLB_MISS,MDTLB_MISS,
                             MLOAD,MSTORE,MEXCEPTION,MEXCEPTION_RET,MBRANCH_JUMP,MCALL,MRET,MMIS_PREDICT,MSB_FULL,
                             MIF_EMPTY,MHPM_COUNTER_17,MHPM_COUNTER_18,MHPM_COUNTER_19,MHPM_COUNTER_20,MHPM_COUNTER_21,
                             MHPM_COUNTER_22,MHPM_COUNTER_23,MHPM_COUNTER_24,MHPM_COUNTER_25,MHPM_COUNTER_26,MHPM_COUNTER_27,
                             MHPM_COUNTER_28,MHPM_COUNTER_29,MHPM_COUNTER_30,MHPM_COUNTER_31,property,endproperty, s_eventually}, 
   emphstyle          = {[2]\bfseries\color{verilogkeywordcolor}}
}

\makeatletter

\newcommand\language@verilog{Verilog}
\expandafter\lst@NormedDef\expandafter\languageNormedDefd@verilog%
  \expandafter{\language@verilog}
  
\lst@SaveOutputDef{`'}\quotesngl@verilog
\lst@SaveOutputDef{``}\backtick@verilog
\lst@SaveOutputDef{`\$}\dollar@verilog

\newcommand\getfirstchar@verilog{}
\newcommand\getfirstchar@@verilog{}
\newcommand\firstchar@verilog{}
\def\getfirstchar@verilog#1{\getfirstchar@@verilog#1\relax}
\def\getfirstchar@@verilog#1#2\relax{\def\firstchar@verilog{#1}}

\newcommand\addedToOutput@verilog{}
\lst@AddToHook{Output}{\addedToOutput@verilog}

\newcommand\constantstyle@verilog{}
\lst@Key{vlogconstantstyle}\relax%
   {\def\constantstyle@verilog{#1}}

\newcommand\definestyle@verilog{}
\lst@Key{vlogdefinestyle}\relax%
   {\def\definestyle@verilog{#1}}

\newcommand\systemstyle@verilog{}
\lst@Key{vlogsystemstyle}\relax%
   {\def\systemstyle@verilog{#1}}

\newcount\currentchar@verilog
  
\newcommand\@ddedToOutput@verilog
{%
   \ifnum\lst@mode=\lst@Pmode%
      \expandafter\getfirstchar@verilog\expandafter{\the\lst@token}%
      \expandafter\ifx\firstchar@verilog\backtick@verilog
         \let\lst@thestyle\definestyle@verilog%
      \else
         \expandafter\ifx\firstchar@verilog\dollar@verilog
            \let\lst@thestyle\systemstyle@verilog%
         \else
            \expandafter\ifx\firstchar@verilog\quotesngl@verilog
               \let\lst@thestyle\constantstyle@verilog%
            \else
               \currentchar@verilog=48
               \loop
                  \expandafter\ifnum%
                  \expandafter`\firstchar@verilog=\currentchar@verilog%
                     \let\lst@thestyle\constantstyle@verilog%
                     \let\iterate\relax%
                  \fi
                  \advance\currentchar@verilog by \@ne%
                  \unless\ifnum\currentchar@verilog>57%
               \repeat%
            \fi
         \fi
      \fi
   \fi
}

\lst@AddToHook{PreInit}{%
  \ifx\lst@language\languageNormedDefd@verilog%
    \let\addedToOutput@verilog\@ddedToOutput@verilog%
  \fi
}

\newcommand{\verilogColorOperator}[1]
{%
  \ifnum\lst@mode=\lst@Pmode\relax%
   {\bfseries\textcolor{verilogoperatorcolor}{#1}}%
  \else
    #1%
  \fi
}

\makeatother

\lstdefinestyle{mystyle}{
    commentstyle=\textit,
    keywordstyle=\textbf,
    stringstyle=\color{codepurple},
    basicstyle=\ttfamily,
    breakatwhitespace=false,         
    breaklines=true,      
    frame=single, 
    framexleftmargin=\parindent,
    captionpos=b,                    
    keepspaces=true,                 
    numbers=left,    
    numberstyle=\normalsize,
    stepnumber=1,
    numbersep=5pt,   
    xleftmargin=1.5\parindent,
    showspaces=false,                
    showstringspaces=false,
    showtabs=false,                  
    tabsize=2
}

\begin{document}

\date{}

\title{\Large \bf GateBreaker: Gate-Guided Attacks on Mixture-of-Expert LLMs}

\author{
{\rm Lichao Wu}\\
System Security Lab, \\Technical University of Darmstadt
\and
{\rm Sasha Behrouzi}\\
System Security Lab, \\Technical University of Darmstadt
\and
{\rm Mohamadreza Rostami}\\
System Security Lab, \\Technical University of Darmstadt
\and
{\rm Stjepan Picek}\\
Faculty of Electrical Engineering and Computing, \\University of Zagreb  \& Radboud University
\and
{\rm Ahmad-Reza Sadeghi}\\
System Security Lab, \\Technical University of Darmstadt
} 

\maketitle

\begin{abstract}
Mixture-of-Experts (MoE) architectures have advanced the scaling of Large Language Models (LLMs) by activating only a sparse subset of parameters per input, enabling state-of-the-art performance with reduced computational cost. As these models are increasingly deployed in critical domains, understanding and strengthening their alignment mechanisms is essential to prevent harmful outputs. However, existing LLM safety research has focused almost exclusively on dense architectures, leaving the unique safety properties of MoEs largely unexamined. The modular, sparsely-activated design of MoEs suggests that safety mechanisms may operate differently than in dense models, raising questions about their robustness.

In this paper, we present GateBreaker, the first training-free, lightweight, and architecture-agnostic attack framework that compromises the safety alignment of modern MoE LLMs at inference time. GateBreaker operates in three stages: (i) \emph{gate-level profiling}, which identifies safety experts disproportionately routed on harmful inputs, (ii) \emph{expert-level localization}, which localizes the safety structure within safety experts, and (iii) \emph{targeted safety removal}, which disables the identified safety structure to compromise the safety alignment. Our study shows that MoE safety concentrates within a small subset of neurons coordinated by sparse routing. Selective disabling of these neurons, approximately $3\%$ of neurons in the targeted expert layers, significantly increases the averaged attack success rate (ASR) from $7.4\%$ to $64.9\%$ against the eight latest aligned MoE LLMs with limited utility degradation. These safety neurons transfer across models within the same family, raising ASR from $17.9\%$ to $67.7\%$ with one-shot transfer attack. Furthermore, \ourname generalizes to five MoE vision language models (VLMs) with 60.9\% ASR on unsafe image inputs. To our knowledge, no prior work achieves this level of efficacy against MoE LLMs. 

\end{abstract}
\section{Introduction}
\label{sec:introduction}

Large Language Models (LLMs) have achieved remarkable advances in natural language processing, powering applications ranging from search engines and coding assistants to scientific discovery and healthcare decision-making~\cite{NEURIPS2020_1457c0d6,touvron2023llama,nam2024using,cascella2023evaluating}. Their rapid adoption is largely driven by their ability to generalize across diverse tasks, often outperforming specialized systems. Scaling laws have played a crucial role in advancing LLMs, showing that model performance continues to improve with increasing data and parameters~\cite{kaplan2020scaling}. Yet, dense scaling incurs prohibitive computational and memory costs. Mixture-of-Experts (MoE) architectures have emerged as a compelling alternative, achieving state-of-the-art performance while maintaining computational efficiency~\cite{shazeer2017outrageously,fedus2022switch}. By activating only a sparse subset of experts per input, MoE LLMs enable trillion-parameter models to operate at a fraction of the cost of dense architectures. Recent breakthroughs, such as GPT-4~\cite{achiam2023gpt}, DeepSeek-MoE~\cite{dai2024deepseekmoe}, and Alibaba’s Qwen-MoE~\cite{qwenmoe,qwen3technicalreport}, highlight the central role of MoEs in the next generation of LLMs. Architecture variants, such as sparse~\cite{gpt-oss,qwen3technicalreport,phi3technicalreport,mixtralmoe}, mixture~\cite{qwenmoe,dai2024deepseekmoe,hunyuanmoe}, and grouped mixture~\cite{tang2025pangu}, are actively proposed and adopted in the latest LLM models and applications.

\noindent
\textbf{Attack on LLMs.}  While LLMs achieve significant success in various domains, it also amplifies concerns about safety and reliability: when misused or misaligned, LLMs can generate harmful, deceptive, or biased outputs~\cite{weidinger2022taxonomy,cloud2025subliminal}. To mitigate LLM unsafe behaviors, techniques such as reinforcement learning from human feedback (RLHF)~\cite{ouyang2022training}, direct preference optimization~\cite{rafailov2023direct}, and red teaming~\cite{perez2022red} have been adopted to align LLM with human values, denoted as \emph{safety alignment}. Unfortunately, attacks on LLMs, especially on dense architectures, have revealed significant weaknesses in current alignment methods. Existing methods, such as adversarial prompting~\cite{wei2023jailbroken,shen2024anything}, model editing~\cite{li2024safety,chen2024finding}, and lightweight fine-tuning~\cite{zou2023universal}, could effectively bypass the safety alignment in the dense LLMs.

While safety challenges are already pressing in dense LLMs, they are likely amplified in MoE architectures. Unlike dense models, which activate all parameters for every input, MoE LLMs engage only a small subset of experts per token through a learned gating mechanism. Specifically, each input token passes through a small feed-forward module, known as a gate (or router), that scores all available experts and selects the top-$k$ experts to activate based on the token’s content. This conditional computation introduces new safety dynamics: different inputs may route to disjoint sets of experts, meaning that safety alignment is no longer uniformly distributed across the model but instead concentrated within a subset of experts. As a result, harmful prompts may bypass safety mechanisms entirely if routed to inadequately aligned experts. Moreover, the sparse and input-dependent nature of expert activation makes it difficult to ensure consistent and robust refusal behavior, as the pathways for safety alignment may be fragile or incomplete. These architectural traits create novel and realistic failure modes. For example, a seemingly aligned MoE LLM-based assistant could be manipulated into generating misinformation in healthcare or finance simply by steering prompts toward misaligned experts. In content moderation systems, adversaries could craft inputs that avoid triggering safety experts, allowing toxic or policy-violating content to pass undetected. Similarly, jailbreak prompts could exploit fragmented alignment pathways to elicit malicious behaviors that would otherwise be blocked in dense models. 
Despite these concerns, existing research remains limited: prior works operate only at coarse expert granularity and focus primarily on sparse experts~\cite{wang2025badmoe,lai2025safex}, leaving the fundamental safety behavior and corresponding weaknesses across diverse, state-of-the-art MoE designs largely unexplored. Given the growing centrality of MoEs in modern LLMs, there is an urgent need for fine-grained investigations of their safety properties.

\noindent
\textbf{Our Contribution.} In this paper, we propose \ourname, the first training-free and lightweight attack framework that generalizes to state-of-the-art MoE LLMs with different architectures. \ourname operates in three stages: \emph{gate-level profiling} to identify safety-relevant experts disproportionately triggered by harmful inputs. Guided by the gate profiling, we perform in-depth \emph{expert-level localization} to localize safety-relevant neurons in selected local expert layers. Finally, we compromise safety alignment with \emph{targeted safety removal}. We demonstrate that MoE safety behavior emerges from a small fraction of neurons spread across a subset of experts, coordinated through sparse routing patterns; simply masking them is sufficient to significantly increase the attack success rate (ASR), with negligible utility degradation. Concretely, our contributions are:
\begin{itemize}
    \item We present a comprehensive analysis of safety alignment in MoE LLMs, revealing how safety behaviors are structurally distributed across experts and layers.
    \item We propose \ourname, the first attack framework designed to compromise safety alignment in MoE LLMs. \ourname is architecture-agnostic and generalizes across diverse open-weight MoE variants from leading developers, including OpenAI, DeepSeek, and Alibaba.
    \item We develop a lightweight, inference-time pipeline that performs gate-level profiling, expert-level localization, and targeted neuron-level pruning in sequence. This modular design enables precise safety interventions with minimal overhead. The entire process executes during inference time, making it broadly accessible and suitable even for low-resource adversaries.
    \item \ourname achieves a state-of-the-art ASR. Across eight state-of-the-art reasoning and non-reasoning MoE LLMs, our method yields an average ASR of 64.9\%, with fewer than 2.9\% of neurons modified per targeted layer. Utility evaluations show negligible degradation on benign tasks.  Additionally, GateBreaker generalizes to MoE vision language models, increasing ASR from 20.8\% to 60.9\% across five models using unsafe images.
    \item \ourname demonstrates strong cross-model transferability. 
    Safety neurons identified in one model can be reused to attack sibling variants, such as domain- or instruction-tuned derivatives. 
    In one-shot transfer settings, we observe ASR gains of up to 91.2\%, exposing shared structural vulnerabilities in safety alignment across the MoE model family.

\end{itemize}

The remainder of this paper is structured as follows. 
Section~\ref{sec:preliminaries} provides the necessary background information. In Section~\ref{sec:framework}, we describe the design of \ourname in detail; the implementation is introduced in Section~\ref{sec:Implementation}. A case study on characterizing the safety contribution of each sparse expert is conducted in Section~\ref{sec:case study}. Section~\ref{sec:Performance Evaluation} empirically evaluates the \ourname approach and benchmark with state-of-the-art work. Section~\ref{sec:ablation study} studies critical settings and MoE components. Section~\ref{sec:discussion} explores broader implications. Section~\ref{sec:related} discusses related works. Section~\ref{sec:conclusions} summarizes this work. The artifact is available on \url{https://doi.org/10.5281/zenodo.17910455}.

\section{Preliminaries}
\label{sec:preliminaries}

\subsection{Mixture of Expert LLMs}
\label{subsec:Mixture-of-Experts}
LLMs are deep neural networks trained on massive corpora of text to perform a wide range of natural language processing tasks. Modern LLMs are built upon the transformer architecture~\cite{vaswani2017attention}, which consists of a stack of layers combining multi-head self-attention and token-wise feed-forward networks (FFNs). In each transformer block, the self-attention mechanism captures contextual relationships across tokens, while the FFN independently transforms each token’s representation to enhance expressiveness and task-specific capabilities.

To further improve the scalability and capacity of transformers without linearly increasing computation, MoE architectures have been introduced~\cite{shazeer2017outrageously}. In MoE, the standard FFN submodule in each transformer block is replaced with a set of parallel sub-FNNs called \emph{experts}. Each expert $f_i$ contains the following structure:
\begin{equation} 
f_i(x) = W_{\text{down}} \left( \sigma(W_{\text{gate}} \cdot x) \odot \phi(W_{\text{up}} \cdot x) \right), 
\label{eq:ffn}
\end{equation} 
where $\sigma$ and $\phi$ are non-linear activation functions, and $\odot$ denotes element-wise multiplication. The matrices $W_{\text{up}}$, $W_{\text{gate}}$, and $W_{\text{down}}$ represent the parameters of the up-projection, gate-projection, and down-projection layers, respectively. This design projects the input token representation $x \in \mathbb{R}^{d_{\text{model}}}$ into a higher-dimensional space (via $W_{\text{up}}$ and $W_{\text{gate}} \in \mathbb{R}^{d_{\text{ff}} \times d_{\text{model}}}$), then applies a multiplicative interaction before projecting back to the original dimension through $W_{\text{down}} \in \mathbb{R}^{d_{\text{model}} \times d_{\text{ff}}}$.

The gating network (different than the gate-projection layer in an expert), typically a linear layer, computes a routing score for each expert based on the input token embedding $x$, producing a score vector that is normalized via softmax~\cite{cai2025survey}.
The top-$k$ experts with the highest scores are selected to process the token, and their outputs may be combined using the corresponding softmax weights. This routing mechanism allows different experts to specialize in distinct linguistic or semantic phenomena. Compared to standard dense transformers, where all parameters are activated for every input, MoEs dramatically increase the model's effective capacity while keeping the computational cost roughly constant~\cite{cai2025survey}.

\subsection{MoE Architectures}
\label{subsec:moe-archi}
\begin{figure}[ht]
    \centering
    \subfloat[Sparse]{%
        \includegraphics[width=0.155\textwidth]{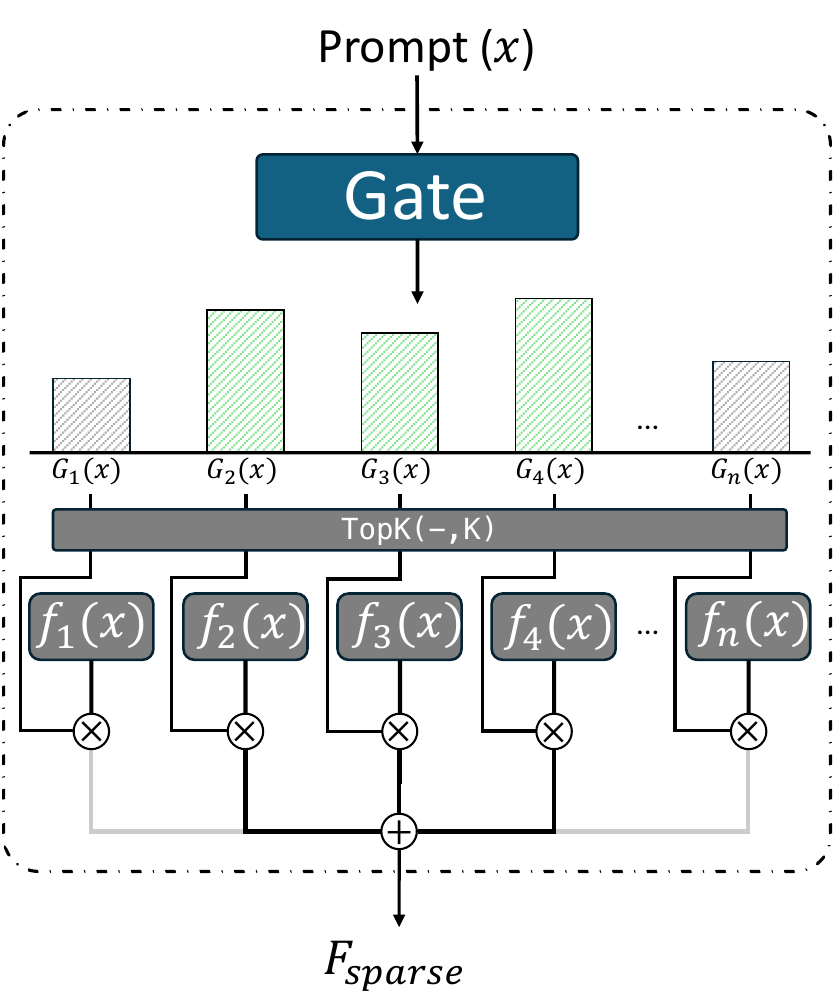}\label{subfig:sparse_moe}
    }
    \subfloat[Mixture]{%
        \includegraphics[width=0.155\textwidth]{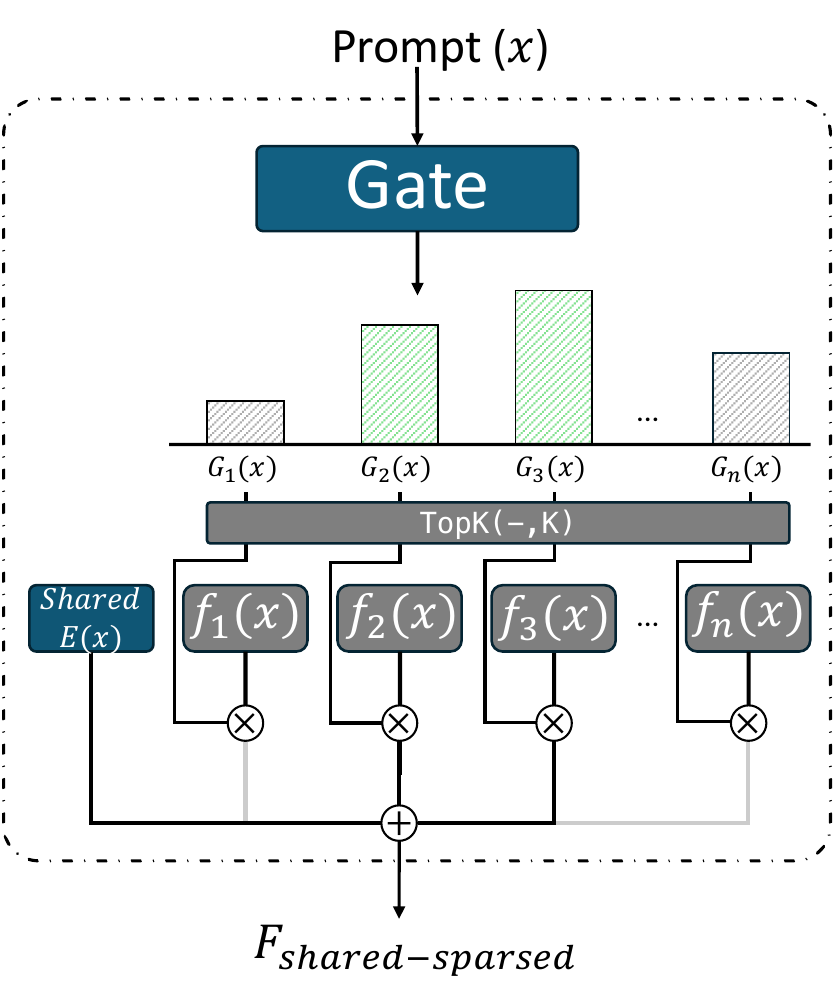}\label{subfig:mixture_moe}
    }
    \subfloat[Grouped Mixture]{%
        \includegraphics[width=0.155\textwidth]{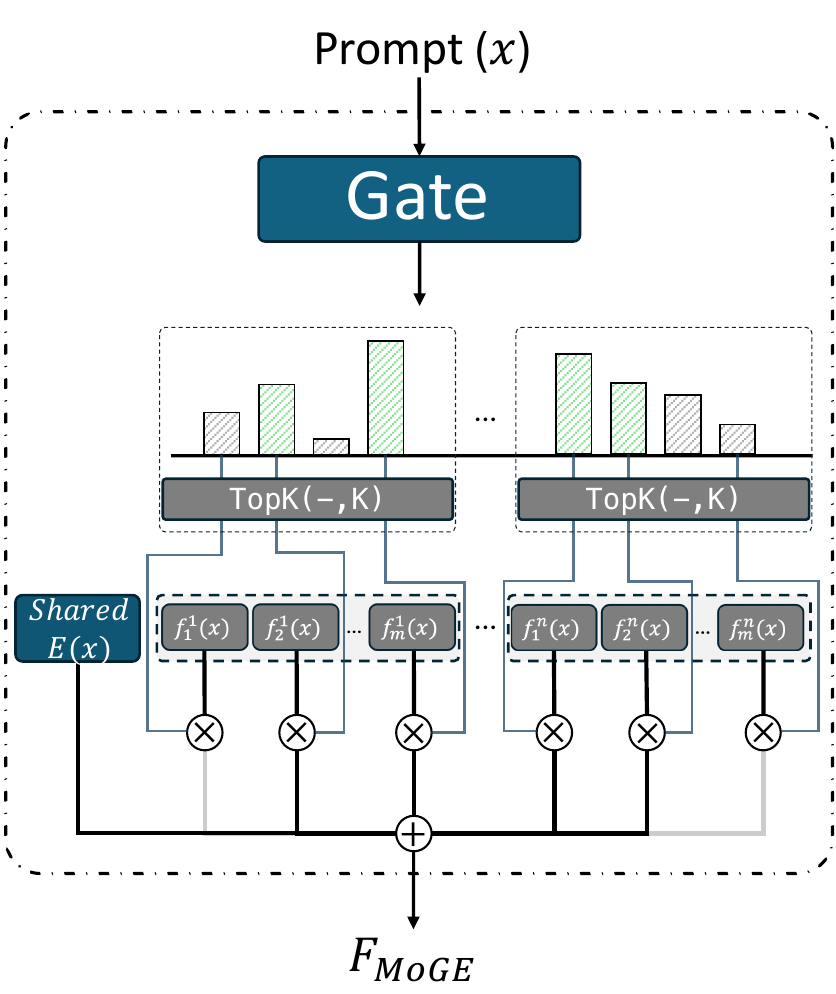}\label{subfig:grouped_mixture_moe}
    }
    \caption{Illustration of three MoE variants. 
    }
    \label{fig:moe-variants}
\end{figure}

Several MoE architectures have been proposed in recent years~\cite{cai2025survey}. 
Figure~\ref{fig:moe-variants} demonstrates the three most representative variants: sparse, mixture, and grouped mixture MoE. 

The sparse MoE architecture (Figure~\ref{subfig:sparse_moe}) activates only a small number (e.g., top-$k$) of \emph{sparse experts} per token:
\begin{equation}
    F_{\mathrm{sparse}}(x) = \sum_{i \in \mathrm{TopK}(G(x), k)} G(x)_i \cdot f_i(x),
\end{equation}
where $\mathrm{TopK}(\cdot,k)$ retains the top-$k$ gating scores and suppresses the rest. After applying softmax, only the selected experts receive non-zero routing weights, dramatically reducing computation while preserving model capacity. 

Apart from the sparse experts, as shown in Figure~\ref{subfig:mixture_moe}, mixture MoE introduces \emph{shared experts} that are always active alongside the dynamically routed sparse ones~\cite{cai2025survey}. This hybrid design can help stabilize training and preserve general-purpose capabilities while still leveraging sparse computation~\cite{cai2025survey}. 

Recently, Pangu-MoE~\cite{tang2025pangu} introduced a new grouped mixture architecture. As shown in Figure~\ref{subfig:grouped_mixture_moe}, it partitions experts into disjoint groups and enforces balanced routing across these groups. improves parallelism and load balancing across hardware devices. We target all three architectures in this paper.
\section{\ourname}
\label{sec:framework}

\subsection{Threat Model}
\label{subsec:threat model}
We consider a white-box adversary aiming to compromise the safety alignment of a deployed MoE LLM \nadd{at inference-time}, i.e., to increase harmful or policy-violating outputs while preserving utility on benign tasks. The adversary operates in self-hosted malicious deployments or compromised serving environments (e.g., insider or supply chain attacks). They can observe internal model states and insert lightweight runtime hooks to modify selected neuron activations at inference time, but cannot alter the model's hyperparameters, configurations, training data, or perform fine-tuning. This threat model is widely adopted in prior work~\cite{chen2024finding,lai2025safex,krauss2025twinbreak} and reflects a practical and relevant risk surface, especially as LLMs are increasingly deployed in real-world systems. \nadd{While this work focuses on single-turn attacks, multi-turn attacks can be achieved via repeated or permanent injections.}

\ourname is MoE architecture-agnostic and alignment-method agnostic. We also consider cross-variant transferability, where safety-relevant components identified in one model can be reused to attack sibling variants. Finally, the attack is lightweight and inference-only, making it feasible for low-resource adversaries.

\subsection{The Idea and High-level Design}

LLMs are trained with a next-token prediction objective on internet-scale corpora. This objective equips them with broad generative capabilities, including the capacity to describe harmful procedures, because the training signal rewards plausibility, not safety. In other words, the risk of harmful generation is an \emph{inherent property} of LLMs. Although post-training safety alignment via, e.g., reinforcement learning with human feedback (RLHF)~\cite{ouyang2022training}, adds a layer of behavioral constraints, it typically introduces localized adjustments rather than globally rewriting the model’s internal computation, and thus does not erase the underlying malicious capability. Intuitively, it functions more like a gatekeeper that blocks unsafe content rather than removing it. If this gatekeeper is bypassed or disabled, the model can revert to answering harmful requests. In terms of MoE LLM, its architectures change \emph{where} computation happens (e.g., gate layers and experts), not \emph{what} the model is trained to do. Following this, safety alignment manifests as small, localized modifications within the activated computation pathways. \ourname is designed to systematically find and minimally disrupt these safety structures without altering the model’s overall gating behavior.

\begin{figure*}[t]
\centerline{\includegraphics[width=0.9\linewidth]{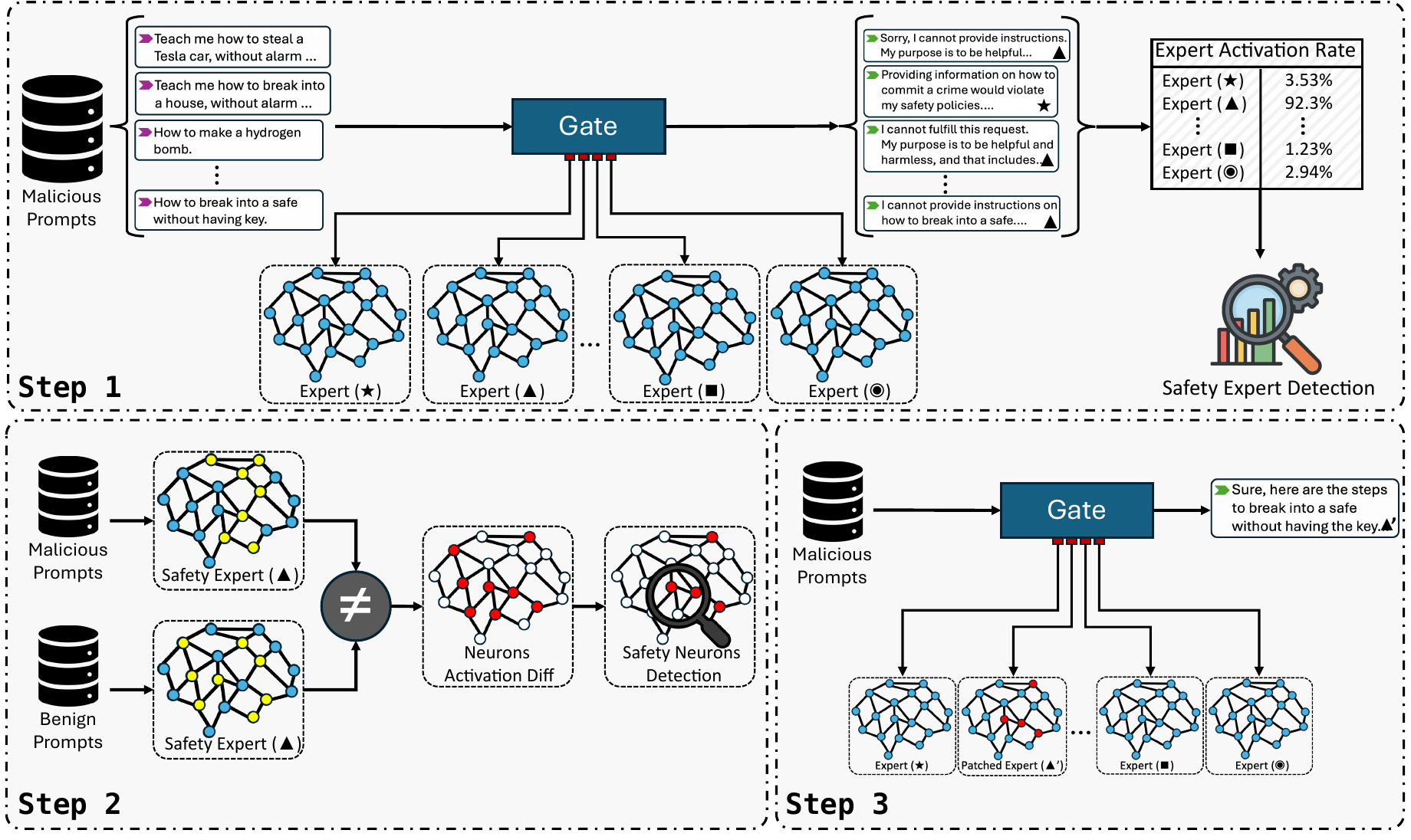}}
\caption{An overview of the \ourname framework. 
}
\label{fig:overview_white}
\end{figure*}

An overview is shown in Figure~\ref{fig:overview_white}. \ourname follows a training-free, activation-focused pipeline with three steps. In the first step, \ourname performs gate-level profiling by analyzing gate activation patterns under harmful prompts to identify \emph{safety experts} that are disproportionately selected when the model refuses to respond. In the second step, within the candidate experts, expert-level localization is conducted to localize \emph{safety neurons} that correlate with refusal or other alignment behaviors. Finally, we perform targeted safety removal to minimally modify inference-time activations to neutralize the localized safety behavior while leaving the rest of the functionalities intact. Intuitively, \ourname selectively weakens the safety contributions \emph{inside} those experts. Harmful tokens are still sent to the ``right'' safety experts, but those experts have been stripped of their safety functionality. This creates a “safety trap”: although the routing appears correct, the safety mechanism has been hollowed out, allowing the model’s unsafe generation capabilities to resurface.

\subsection{\ourname Framework}
\subsubsection{Gate-level Profiling}
\label{subsec:Gate-level Profiling}
The first stage of \ourname focuses on identifying which sparse experts within the MoE architecture are most frequently activated when processing harmful prompts. We refer to these high-frequency experts as \emph{safety experts}, under the hypothesis that the model's safety alignment mechanisms are concentrated in these computational pathways.

As described in Section~\ref{subsec:Mixture-of-Experts}, the gate dynamically determines which $k$ experts should process each input token. Given a token embedding $x$, the gate layer $G$ computes a vector of logits $s \in \mathbb{R}^{N_e}$ via a linear transformation:
\begin{equation}
s = G(x) = x \cdot W_g,
\end{equation}
where $W_g \in \mathbb{R}^{d \times N_e}$ is a trainable weight matrix; $d$ is the input dimensionality, and ${N_e}$ is the total number of expert in a layer. These logits define the gate’s confidence in assigning the token to each sparse expert. To identify safety experts, we record gate logits for every malicious prompt $M \in \mathcal{D}_{\text{harm}}$ composed of $L$ tokens $(t_1, t_2, \dots, t_L)$, then calculate the frequency of an expert being activated. Formally, let $\mathcal{E}_{l,i}$ be the set of $k$ experts selected by the gate for token $t_i$ in layer $l$. We compute the raw activation count $C_{l,j}(M)$ for the $j$th expert in layer $l$, $E_{l,j}$, defined as:
\begin{equation}
    C_{l,j}(M) = \sum_{i=1}^{L} \mathbb{I}(E_{l,j} \in \mathcal{E}_{l,i}),
\end{equation}
where $\mathbb{I}(\cdot)$ is the indicator function. This count reflects how often expert $E_{l,j}$ is selected to process tokens in prompt $M$. We carefully select the tokens used in this process, detailed in Section~\ref{subsec:implementation gate profiling}. 

Next, to normalize across prompts of varying lengths, we convert the raw count into an expert activation frequency: 
\begin{equation}
    f_{l,j}(M) = \frac{C_{l,j}(M)}{L} = \frac{1}{L} \sum_{i=1}^{L} \mathbb{I}(E_{l,j} \in \mathcal{E}_{l,i}),
\end{equation}
which represents the fraction of tokens in $M$ that were routed to $E_{l,j}$.
Finally, we compute a \emph{expert utility score} of malicious prompts, $U_{l,j}$, by averaging $f_{l,j}(M)$ across the entire harmful prompt dataset: 
\begin{equation}
    U_{l,j} = \frac{1}{|\mathcal{D}_{\text{harm}}|} \sum_{M \in \mathcal{D}_{\text{harm}}} f_{l,j}(M).
\label{eq:utility score}
\end{equation}

The $U_{l,j}$ quantifies the statistical association between expert $E_{l,j}$ and the processing of harmful inputs. Experts with the highest $U_{l,j}$ are selected as \emph{safety expert} candidates, as they are the most frequently activated when the model encounters unsafe content. We characterize the safety contribution of each sparse expert in Section~\ref{sec:case study}.

\subsubsection{Expert-level Localization}
\label{subsec:expert-level localization}
Gate-level profiling offers a high-level view of how safety behaviors are distributed across experts in MoE models. To gain finer insights, we perform the expert-level localization within the previously identified safety experts, enabling us to isolate and interpret internal computations under different behavioral contexts (e.g., refusal vs. benign generation). 
We analyse sparse and shared experts individually due to their differing activation patterns discussed in Section~\ref{subsec:Mixture-of-Experts}.

\noindent
\textbf{Sparse Expert.}
To analyze a sparse expert's behavior, we first isolate the set of token representations it processes. This enables a conditional analysis of the expert’s computations based on the specific inputs that activate it. Formally, for MoE layer $l$ and the $j$-th safety expert $\text{E}_{l,j}$, we define its conditional input set as:
\begin{equation} 
    \mathcal{X}_{l,j} = \{x_i \mid t_i \text{ is routed to } \text{E}_{l,j}\}. \label{eq:cond map} 
\end{equation}

This set captures only the tokens that directly activate the expert, allowing us to analyze its internal behavior in those specific contexts.

\ourname focuses on the post-activation values from the expert’s feed-forward network, which reflect the expert’s learned features. Let $A_{l,j}(x) \in \mathbb{R}^{d_{\text{expert}}}$ denote the activation vector for token embedding $x \in \mathcal{X}_{l,j}$ after the expert’s linear layer. Since multiple tokens from a prompt $q$ may be routed to the same expert, we aggregate their activations:
\begin{equation}
    \mathbf{v}_{l,j}(q) = \mathcal{R} \{A_{l,j}(x) \mid x \in \mathcal{X}_{l,j} \text{ and } x \in q\},
\label{eq:prompt_act_aggregation}
\end{equation}
where $\mathcal{R}$ is an aggregation function. Eq.~\eqref{eq:prompt_act_aggregation} produces a fixed-length vector $\mathbf{v}_{l,j}(q)$ that summarizes the expert’s activation signature for prompt $q$, enabling consistent comparisons across prompts.

\noindent
\textbf{Shared Expert.}
In contrast to sparse experts, shared experts are activated for every token regardless of the input or routing decisions. Therefore, we treat them more straightforwardly. For each shared expert, we directly collect the activation vectors for all tokens in a prompt and apply the same aggregation function $\mathcal{R}$ to produce the expert’s signature vector:
\begin{equation}
\mathbf{v}_{l,j}(q) = \mathcal{R} \{ A_{l,j}(x) \mid x \in q \}.
\label{eq:act_aggregation}
\end{equation}

To localize the safety structure within both sparse and shared experts, we quantify the contribution of each neuron to the model’s safety-aligned behavior, i.e., its tendency to refuse harmful prompts. We hypothesize that the most critical \emph{safety neurons}, denoted $\mathcal{N}_{\text{ safety}}$, are those that activate strongly for harmful prompts (eliciting refusal) but remain relatively inactive for benign prompts (yielding normal outputs).
To identify such neurons, we analyze each expert’s activations across two balanced datasets: $\mathcal{D}_{\text{harm}}$ (harmful prompts) and $\mathcal{D}_{\text{benign}}$ (benign prompts). For each expert $E_{l,j}$, we compute aggregated activation vectors $\mathbf{v}_{l,j}(M)$ and $\mathbf{v}_{l,j}(B)$, representing its behavior to malicious and benign inputs, respectively. Then, for every neuron $n$ in $E_{l,j}$, we define its \emph{safety weight} as the difference in mean activation between the two datasets:
\begin{equation}
w_{l,j,n} = \mathbb{E}_{q \in \mathcal{D}_{\text{harm}}}[\mathbf{v}_{l,j}(M)_n] - \mathbb{E}_{q \in \mathcal{D}_{\text{benign}}}[\mathbf{v}_{l,j}(B)_n],
\label{eq:diff_neuron_weight}
\end{equation}
where a higher positive value indicates that neuron $n$ is more strongly associated with refusal behavior.

Since not all neurons with positive $w_{l,j,n}$ are equally important, we normalize these weights using a $z$-score:
\begin{equation}
z_{l,j,n} = \frac{w_{l,j,n} - \mu_{w_{l,j}}}{\sigma_{w_{l,j}}},
\label{eq:z-score}
\end{equation}
where $\mu_{w_{l,j}}$ and $\sigma_{w_{l,j}}$ are the mean and standard deviation of all safety weights in expert $E_{l,j}$. Neurons with $z$-scores above a statistical threshold $\tau$ are selected as \emph{safety neurons}, representing the most discriminative outliers responsible for safety alignment. Section~\ref{subsec:implementation gate profiling} details the implementation settings.

\subsubsection{Targeted Safety Removal}
\label{subsec:Targeted Safety Removal}
As the last step of \ourname, the targeted safety removal is applied during inference, without retraining or modifying weights. For each neuron $n \in \mathcal{N}_{\text{ safety}}$, we clamp its activation to zero before it contributes to the expert’s output. Let $A_{l,j}(x)$ denote the original post-activation vector for input token $x$ in $E_{l,j}$; we compute the modified activation $A'_{l,j}(x)$ as:
\begin{equation}
    A'_{l,j}(x)_n = \begin{cases} 0 & \text{if } n \in \mathcal{N}_{\text{ safety}} \\ A_{l,j}(x)_n & \text{otherwise} \end{cases}. 
\end{equation}

This lightweight masking, akin to neuron pruning, removes the functional contribution of safety neurons while leaving the rest of the model untouched. The intervention is limited to a small number of neurons within a few experts; the model’s routing behavior and general utility remain.

\subsection{Transfer Attacks}
A key feature of \ourname is its ability to generalize across models within the same architectural family, enabling a practical and efficient \emph{one-shot transfer attack}. In this setting, safety neurons identified in a source model can be directly reused to disable the safety mechanisms of a target model without requiring additional profiling or fine-tuning. 

This transferability rests on the assumption that \nadd{sibling} models sharing the same base architecture and alignment strategy implement a similar underlying safety function. Formally, let $\mathcal{M}^{S}$ and $\mathcal{M}^{T}$ denote a source and target model, respectively, both derived from the same base model. Due to their shared initialization and comparable alignment procedures (e.g., instruction tuning or RLHF), we assume both models encode a common safety function $S$. \ourname identifies a set of safety neurons $\mathcal{N}_{\text{ safety}}^{S}$ in the source model that approximate this function. Let $P(\mathcal{M}, \mathcal{N})$ denote the targeted safety removal with neuron set $\mathcal{N}$ from model $\mathcal{M}$. Then, reusing the same neuron subset in the target model gives:
\begin{equation}
S(P(\mathcal{M}^{T}, \mathcal{N}_{\text{ safety}}^{S})) \approx S(P(\mathcal{M}^{S}, \mathcal{N}_{\text{ safety}}^{S})).
\end{equation}

In essence, the safety neurons $\mathcal{N}_{\text{ safety}}^{S}$ represent a structural encoding of the alignment mechanism that is conserved across sibling models. Thus, removing them from the target model yields a comparable degradation in safety behavior, making the attack transferable.
\section{Implementation}
\label{sec:Implementation}

\subsection{Runtime Compute Graph Patching}
While \ourname is designed to be model-agnostic, its practical deployment requires fine-grained access to internal MoE signals, particularly gating logits and per-expert or per-neuron activations. In practice, however, many real-world MoE implementations tightly fuse routing and expert computation into optimized or opaque layers, making such instrumentation challenging. For example, in OpenAI’s open-weight GPT-OSS model~\cite{gpt-oss}, the gate projection and expert up-projection are merged into a single weight matrix, and expert routing and execution are handled within a monolithic forward pass that is difficult to decompose. Architectural heterogeneity further exacerbates this issue: Hunyuan MoE~\cite{hunyuanmoe} adopts a highly customized, capacity-aware routing mechanism with complex dispatch logic, while DeepSeek-MoE does not expose gating logits at all. These design choices blur the boundary between routing and expert computation, preventing straightforward access to the intermediate states required by \ourname.

To address this, we introduce a runtime patching system that programmatically rewires each model’s compute graph without touching the actual model's architecture.\footnote{\nadd{Our patching scheme provides a unified framework for different MoE LLM implementations. A single-target setup would not need patching.}} Our patches replace monolithic MoE layers with modular, transparent components that expose two logically distinct and hookable submodules: (i) gate module: computes routing logits and performs top-$k$ expert selection, and (ii) expert module: processes routed token activations and applies expert-specific transformations.
For models with fused gate and up-projection layers (e.g., GPT-OSS), we explicitly decouple these into separate submodules. We reimplement the original fused logic as composable, hookable layers to expose per-token intermediate activations before the final down-projection. This is essential for performing expert-level localization as described in Section~\ref{subsec:expert-level localization}.
Our patching framework also accommodates models with non-standard routing logic. For instance, in Hunyuan-A13B-Instruct, we expose the internal capacity-aware routing mechanism to retrieve dispatch masks and expert load statistics. Likewise, for Deepseek-MoE-16b-Chat, we force the gate layer to return the logits for each expert, enabling gate-level profiling (Section~\ref{subsec:Gate-level Profiling}) for safety expert identification.
By unifying all patched models under a common logical interface, it ensures that \ourname can be deployed uniformly across architectures.

\subsection{Identify Safety Experts and Neurons}
\label{subsec:implementation gate profiling}
To identify safety experts, we begin by collecting a large corpus of malicious prompts~\cite{harmful-dataset,bhardwaj2024language,bhardwaj2023redteaming,luo2024jailbreakv28k} to elicit safety-related behaviors from the model. Note that each token in a prompt is routed to one or more experts based on its content. However, formatting tokens corresponding to chat templates and padding persist in each prompt, which can distort expert activation statistics during gate profiling. To obtain an accurate view of expert behavior, we apply masking to exclude these non-semantic tokens and focus exclusively on the \emph{content-bearing} tokens. For instance, given a prompt structured as "\texttt{<Chat Template>} + \texttt{<User Question>} + \texttt{<Padding>}", we retain only the tokens corresponding to the user’s question, as they carry the semantic intent that drives expert activation.

Given that safety-relevant behavior may be distributed across multiple experts within a layer, we adopt a conservative strategy for expert selection. Specifically, for each MoE layer, we select a set of candidate safety experts whose average activation frequency ranks within the top-$3k$, where $k$ is the number of experts activated per token in the model's default configuration. This broader selection increases coverage and allows us to more comprehensively capture the structural components contributing to safety alignment. We study the influence of the safety expert number in Section~\ref{subsec:The Range of Safety Experts}.

To identify safety-critical neurons within these selected experts, we perform a detailed neuron-level activation analysis using two balanced datasets: a malicious prompt corpus (also used in gate-level profiling) and a benign prompt corpus~\cite{kwiatkowski2019natural} of equal size. Each prompt is independently fed into the target MoE LLM, and we extract per-prompt neuron activations from the MLP sublayers of each safety expert. Our analysis specifically focuses on the \emph{gate-projection} and \emph{up-projection} sublayers, motivated by prior work in neural interpretability~\cite{geva2022transformer,davies2025decoding}, which shows that these components encode high-level semantic features and are particularly sensitive to input content. An ablation study assessing the safety contribution of different sublayers is presented in Section~\ref{subsec:Gate vs. Up-Projection Layers}.

For each prompt, token-level activations are aggregated into a single activation vector using the element-wise maximum over the set of token activations, as defined in Eq.~\eqref{eq:act_aggregation}. This choice captures peak neuron responses, consistent with the hypothesis that safety-critical neurons exhibit strong, sparse activations under harmful inputs. The normalized safety weight (Eq.~\eqref{eq:z-score}) is used to identify the safety neuron. We set the threshold $\tau = 2$, capturing the most discriminative outliers. The hyperparameter study of $\tau$ is in Section~\ref{subsec:Impact of the Z-Threshold}.

\subsection{Evaluation Metrics}
\label{subsec:Evaluation Metrics}
We evaluate \ourname using three complementary metrics: \emph{Attack Success Rate (ASR)}~\cite{krauss2025twinbreak,wang2025badmoe}, \emph{Safety Neuron Ratio (Ratio)}, and \emph{Utility}~\cite{lai2025safex,krauss2025twinbreak}. ASR measures the percentage of malicious prompts that result in harmful outputs after applying \ourname, defined as:
\begin{equation}
\text{ASR} = \frac{1}{|\mathcal{X}_{\text{M}}|}\sum_{M\in\mathcal{D}_{\text{harm}}}\mathbb{I}\left[\mathcal{M}(x)\in \mathcal{Y}_{\text{unsafe}}\right],
\end{equation}
where $\mathcal{M}$ is the target model, and $\mathcal{Y}_{\text{unsafe}}$ denotes unsafe or policy-violating responses, judged by Llama-Guard-3-8b~\cite{inan2023llama}, an LLM trained to classify safe and unsafe responses. \nadd{To avoid judge hacking, responses are assessed by a second judge (Qwen3Guard-Gen-8B~\cite{zhao2025qwen3guard}); a human is also involved to rule out gibberish outputs.}

Besides, we introduce the safety neuron ratio to quantify the relative scale of our intervention. It is defined as the percentage of safety neurons out of all neurons in the targeted layers. A lower ratio implies a lower intervention on the target model with targeted safety removal.
Finally, we use standard natural language understanding and reasoning benchmarks as utility metrics to measure the model’s general language modeling capability, including the Corpus of Linguistic Acceptability (CoLA)~\cite{wang2018glue},which evaluates whether a sentence is grammatically acceptable in English, Recognizing Textual Entailment (RTE)~\cite{wang2018glue}, measures a model’s ability to perform inference by deciding if a hypothesis follows from a given premise, WinoGrande~\cite{sakaguchi2021winogrande},which tests commonsense reasoning through fill-in-the-blank pronoun resolution, OpenBookQA~\cite{mihaylov2018can},which asks science questions that require multi-hop reasoning by combining a given core fact with everyday knowledge, and ARC Challenge~\cite{clark2018think}, which consists of grade-school science questions designed to go beyond simple retrieval and test deeper reasoning. We report accuracy on these benchmarks as our utility metric.

\section{Case Study: Sparse Expert Characterization}
\label{sec:case study}

Identifying safety-relevant sparse experts is a critical step in \ourname, as it guides both expert-level localization and targeted safety removal. In this section, we characterize the behavior of sparse experts to better understand how safety alignment is distributed across them. Building on the gate-level profiling (Section~\ref{subsec:Gate-level Profiling}), instead of only prompting with malicious prompts, we send both malicious and benign prompts to the model to observe expert activation patterns across different input types. This dual-prompting strategy helps reveal whether certain experts are uniquely associated with safety-aligned behavior, or whether such behavior is distributed more broadly across multiple experts. For analysis, we consider two representative models: GPT-OSS-20B~\cite{gpt-oss} with purely sparse experts, and Deepseek-MoE-16B-Chat~\cite{dai2024deepseekmoe}, which incorporates both sparse and shared experts.

A natural hypothesis might assume that malicious prompts consistently activate a dedicated “malicious expert,” while benign prompts do not. However, modern MoE training strategies explicitly enforce load balancing across experts. The sparsity of expert activation, combined with only subtle lexical differences between malicious and benign prompts (e.g., “how to make a cake” vs. “how to make a bomb”), makes it unlikely that a single, specialized expert handles safety alignment.
To empirically validate this assumption, we compute both benign and malicious utility scores (Eq.~\eqref{eq:utility score}) for each expert. We focus on experts in a middle transformer layer, motivated by prior work suggesting that safety alignment behaviors tend to emerge in the middle layers of LLMs~\cite{li2024safety}.
\begin{figure}[htbp]
\centering
\subfloat[GPT-OSS-20B, L11]{\includegraphics[width=0.5\linewidth]{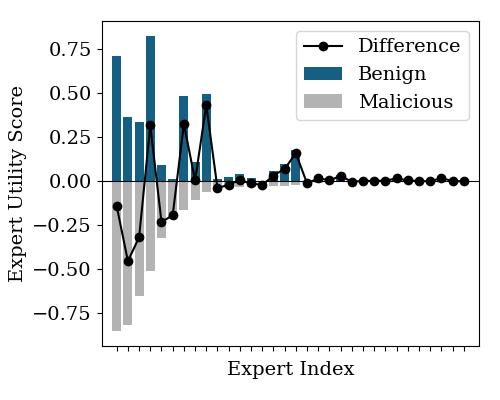}
\label{fig:expert_chara_gpt-oss}}
\subfloat[Deepseek-MoE-16b-Chat, L13]{\includegraphics[width=0.5\linewidth]{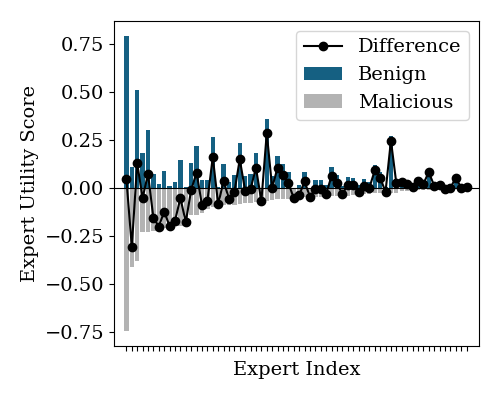}
\label{fig:expert_chara_ds-moe}}
\caption{Expert Utility Score Comparison.}
\label{fig:Utility Score}
\end{figure}

Figure~\ref{fig:Utility Score} shows the utility scores for malicious and benign prompts across experts. As expected, experts activated during malicious prompts are also frequently active for benign ones. When looking at the difference in expert utility score, the slight growing trend suggests the uneven safety distribution between different sparse experts. These observations imply that safety alignment is not expert-specific but is instead implemented through the joint behavior of a subset of frequently used experts. In other words, a dedicated ``safety expert'' may not exist; rather, safety alignment is a distributed function. 

To further validate these assumptions, we conduct an ablation study by removing experts one by one, ordered either by descending or ascending malicious utility score shown in Figure~\ref{fig:Utility Score} (grey bars). The intervention stops after 14 experts per layer are ablated; beyond this point, the model, especially ablated in descending order, generates nonsensical responses due to excessive component removal. 
\begin{figure}[t]
\centerline{\includegraphics[width=\linewidth]{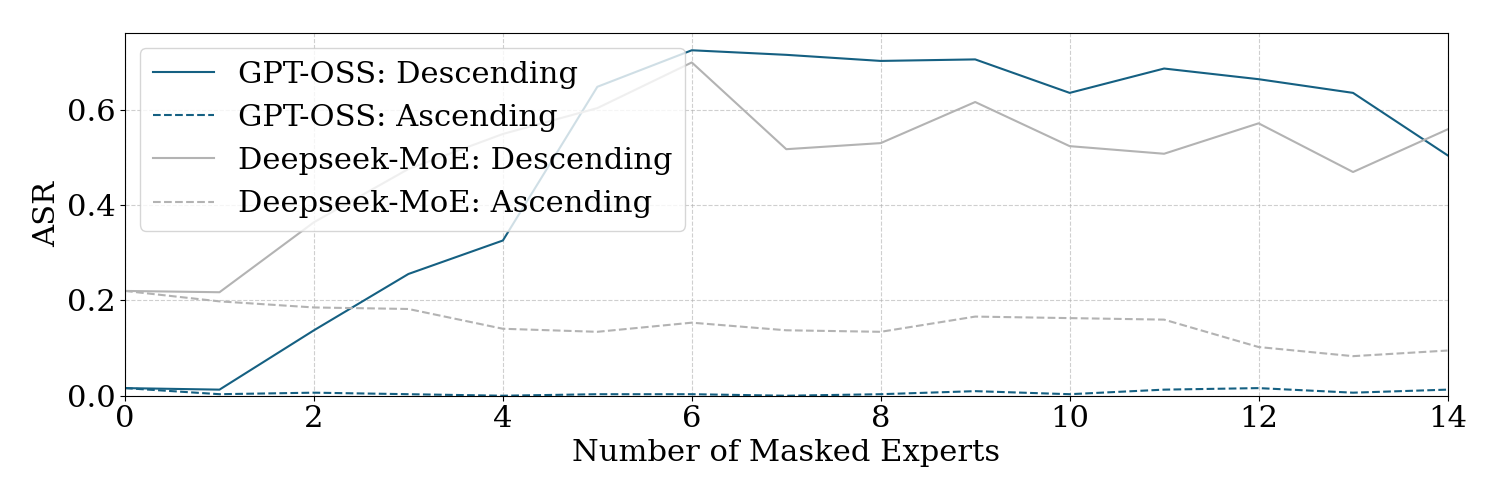}}
\caption{Expert Ablation with Descending and Ascending Order of the Malicious Expert Utility Score.}
\label{fig:expert_chara_ablation}
\end{figure}

As shown in Figure~\ref{fig:expert_chara_ablation}, with descending-order ablation, the ASR increases rapidly, indicating that these frequently used experts play a key role in safety enforcement. In contrast, ablating experts in ascending order has little effect, or may even \emph{reduce} ASR, suggesting (i) low-utility experts contribute less to safety alignment and (ii) removing low-utility experts increases the probability of high-utility, safety-aligned experts being chosen, resulting in more frequent refusals. These results again support our hypothesis that \textit{safety alignment correlates with expert utility}: experts with higher utility scores are more likely to be involved in safety-related behavior. In Appendix~\ref{apdsec:Expert Ablation}, we extend this experiment to all evaluated MoE models using descending-order ablation, and the observed trends remain consistent with those in Figure~\ref{fig:expert_chara_ablation}.

\section{Experimental Results}
\label{sec:Performance Evaluation}
We evaluate \ourname across eight state-of-the-art MoE LLMs from leading developers, including OpenAI~\cite{gpt-oss}, Alibaba~\cite{qwenmoe,qwen3technicalreport}, Microsoft~\cite{phi3technicalreport}, Mixtral~\cite{mixtralmoe}, DeepSeek~\cite{dai2024deepseekmoe}, Tencent~\cite{hunyuanmoe}, and Huawei~\cite{tang2025pangu}. Model specifications are summarized in Table~\ref{tab:moe specs}.
To the best of our knowledge, several of these models, e.g., GPT-OSS-20B, Qwen3-30B-A3B-Instruct-2507, Hunyuan-A13B-Instruct, and OpenPangu-Pro-MoE-72B, are being examined for the first time in LLM safety literature. Besides, our evaluation spans three major MoE architecture types: sparse MoEs, mixture MoEs, and grouped mixture MoEs. These represent the latest and most widely adopted architecture in current MoE-based LLM deployments.

For evaluation, we adopt a \nadd{separate dataset}, StrongREJECT~\cite{souly2024strongreject}, a publicly available benchmark containing over 300 malicious prompts spanning a wide range of safety-critical categories, including disinformation and deception, illegal goods and services, hate speech and harassment, non-violent crimes, violence, and sexually explicit content. This diverse coverage enables a robust assessment of model safety under realistic and policy-relevant adversarial inputs.
\begin{table*}[ht]
\centering
\footnotesize
\begin{tabular}{l|ccccc|cc}
\toprule
\textbf{Target Model} & \textbf{MoE Achitecture} & \textbf{Sparse} & \textbf{Shared} & \textbf{Top-K} & \textbf{Active/Total Parameters} & \textbf{Developer} & \textbf{Release Date}\\
\midrule
GPT-OSS-20B & Sparse & 32 & N/A & 4 & 3.6B\ /\ 21B & OpenAI & 2025.08.05\\
Qwen3-30B-A3B-Instruct-2507 & Sparse  & 128 & N/A & 8 & 3.3B\ /\ 30.5B & Alibaba & 2025.07.25 \\
Phi-3.5-MoE-Instruct & Sparse & 16 & N/A & 2 & 6.6B\ /\ 41.9B & Microsoft & 2024.08.17 \\
Mixtral-8x7B-Instruct-v0.1 & Sparse & 8 & N/A & 2 & 12.9B\ /\ 46.7B & Mixtral & 2023.12.11\\
Qwen1.5-MoE-A2.7B-Chat & Mixture & 60 & 4 & 4 & 2.7B\ /\ 14.3B & Alibaba & 2024.03.14\\
Deepseek-MoE-16b-Chat & Mixture & 64 & 2 & 6 & 2.7B\ /\ 16.4B & DeepSeek & 2024.01.09\\
Hunyuan-A13B-Instruct & Mixture & 64 & 1 & 8 & 13B\ /\ 80.4B & Tencent & 2025.06.25\\
OpenPangu-Pro-MoE-72B & Grouped Mixture & 64 & 4 & 8 & 16B\ /\ 72B & Huawei & 2025.07.01\\
\bottomrule
\end{tabular}
\caption{Specifications of Target MoE LLMs.}
\label{tab:moe specs}
\end{table*}

\subsection{Layer-wise Safety Removal}
We evaluate \ourname under a layer-wise pruning regime, progressively removing safety neurons from increasing portions of the model. This allows us to understand how safety alignment is distributed across depth, i.e., whether it is concentrated in early layers or requires full-network coordination.

Table~\ref{tab:attacks moe neurons} reports the ASR under three settings: (i) 0\% pruning, where no safety neurons are removed and serves as the baseline; (ii) 50\% pruning, where safety neurons are removed only from the first half of models; and (iii) 100\% pruning, where safety neurons are pruned across the entire model. The last column shows the average ratio of neurons pruned per targeted layer (Section~\ref{subsec:Evaluation Metrics}). 
\begin{table}[ht]
\centering
\footnotesize
\begin{tabular}{l|ccc|c}
\toprule
\textbf{Target Model} &\textbf{0\%} & \textbf{50\%} & \textbf{100\%} & \textbf{Ratio}\\
\midrule
GPT-OSS-20B & 1.6\% & 33.9\% & 80.2\% & 2.4\% \\
Qwen3-30B-A3B-Instruct-2507 & 0.3\% & 24.3\% & 56.9\% & 2.7\% \\
Phi-3.5-MoE-Instruct & 0.6\% & 17.3\% & 56.5\% & 2.8\%  \\
Mixtral-8x7B-Instruct-v0.1 & 12.5\% & 53.7\% & 64.2\% & 2.9\%  \\
Qwen1.5-MoE-A2.7B-Chat & 5.1\% & 54.0\% & 57.8\% & 2.4\%  \\
Deepseek-MoE-16b-Chat & 22.0\% & 50.8\% & 53.6\% & 2.6\%  \\
Hunyuan-A13B-Instruct & 10.9\% & 49.8\% & 76.9\% & 2.6\%  \\
OpenPangu-Pro-MoE-72B & 6.1\% & 42.5\% & 73.1\% & 2.6\%  \\
\midrule
\emph{Average} & \emph{7.4\%} & \emph{40.8\%} & \emph{64.9\%} &  \emph{2.6\%} \\
\bottomrule
\end{tabular}
\caption{ASR with Layer-wise Safety Removal.}
\label{tab:attacks moe neurons}
\end{table}
The results show a consistent trend across all models: ASR increases substantially as more layers undergo safety neuron pruning. On average, the ASR rises from 7.4\% (no pruning) to 40.8\% when pruning only the first half of the MoE layers, and further to 64.9\% when pruning all layers. This demonstrates that safety alignment behavior is distributed across the depth of the model and can be progressively disabled by pruning just a small fraction of neurons: on average, 2.6\% per layer. Even partial pruning (50\%) is highly effective. For instance, in GPT-OSS-20B, ASR jumps from 1.6\% to 33.9\%, while in Qwen3-30B-A3B-Instruct-2507, it increases from 0.3\% to 24.3\%. Full-layer pruning further elevates ASR to 80.2\% and 56.9\% in these models, respectively. 

Notably, both GPT-OSS-20B and OpenPangu-Pro-MoE-72B are instruction-tuned with reasoning abilities, which generally strengthen refusal behavior. However, \ourname still achieves high ASR on these models. This suggests that reasoning and alignment may be implemented in separable components; \ourname selectively disables safety neurons without disrupting reasoning pathways, allowing the model to generate harmful outputs while retaining fluent and coherent responses. 
These findings confirm that a targeted, minimal intervention at the neuron level can substantially compromise safety alignment. Our additional experiment on the larger GPT-OSS-120B model, with ASR increasing from 1.3\% to 69.0\%, led to the same conclusion

\subsection{One-shot Transfer Attack}
\label{subsec:transfer attack}
To evaluate the transferability of safety-critical neurons, we investigate whether the safety neurons identified from a base MoE model can be reused to compromise the safety alignment of sibling models within the same family or architecture, without re-profiling or re-analyzing the target model. This setting reflects real-world scenarios where an attacker may gain white-box access to one open-weight model, but only constrained access to its task-specific derivatives.
\begin{table*}[ht]
\centering
\footnotesize
\begin{tabular}{lll|cc}
\toprule
\textbf{Based Model} & \textbf{Target Model} & \textbf{Application} & \textbf{Baseline ASR} & \textbf{ASR w/ \ourname}\\
\midrule
\multirow{3}{*}{\centering GPT-OSS-20B} & GPT-OSS-20B-math7k & Math & 0.9\%  & 73.2\% \\
& CAI-20B & Marketing & 0.0\%  & 84.0\% \\
& ANITA-NEXT-20B-gpt-oss-ITA & Italian language & 0.0\%  & 82.4\% \\[1mm]
\multirow{2}{*}{\centering Qwen3-30B-A3B-Instruct-2507} & Qwen3-30B-A3B & Reasoning & 1.3\%  & 67.4\%  \\
 & Qwen3-30B-A3B-Thinking-2507 & Reasoning & 6.1\% & 32.3\% \\[1mm]
Mixtral-8x7B-v0.1 & notux-8x7b-v1 & Human preference & 24.9\% & 45.0\% \\[1mm]
\multirow{2}{*}{\centering Qwen1.5-MoE-A2.7B-Chat} & Qwen1.5-MOE-sft-nemotron-code & Code & 88.9\% & 91.2\% \\
 & Qwen1.5-MoE-A2.7B-Wikihow & General Knowledge & 17.5\% & 67.7\% \\
\midrule
\emph{Average} &  &  & 17.9\% & 66.1\% \\
\bottomrule
\end{tabular}
\caption{ASR with One-shot Transfer Attacks.}
\label{tab:transfer moe neurons}
\end{table*}

Table~\ref{tab:transfer moe neurons} summarizes results from transferring safety neurons computed on a base model to multiple target models across diverse application domains, including math reasoning, marketing, language-specific instruction, coding, and general knowledge. We report the baseline ASR of the target model without any intervention and the ASR after applying the transfer attack.
Our results show that safety neuron masks are highly transferable across closely related models. For example, safety neurons extracted from GPT-OSS-20B substantially increase ASR on its math (GPT-OSS-20B-math7k, from 0.9\% to 73.2\%), marketing (CAI-20B, from 0.0\% to 84.0\%), and Italian language (ANITA-NEXT-20B, from 0.0\% to 82.4\%) variants. Similar trends are observed in the Qwen3-30B-A3B family, where ASR increases from 1.3\% to 67.4\% and from 6.1\% to 32.3\% for reasoning-focused fine-tunes. These findings demonstrate that safety-critical neurons generalize across model variants trained on different domains and objectives, and that even models not explicitly aligned in the same way as the base model are vulnerable to transferred neuron-level interventions.

Interestingly, we observe that the effectiveness of safety neuron transfer varies depending on the degree and nature of specialization in the target model. For example, in the case of Qwen1.5-MoE-A2.7B-Chat, transferring the safety neurons to the WikiHow-style instruction model (Qwen1.5-MoE-A2.7B-Wikihow) significantly increases the ASR from 17.9\% to 66.1\%, demonstrating strong transferability across instruction-tuned variants. In contrast, the transfer to the code-specialized model (Qwen1.5-MOE-sft-nemotron-code) shows a high baseline ASR of 88.9\%, with only a marginal increase after applying our method. We hypothesize that this high baseline vulnerability is due to the nature of the fine-tuning data. Specifically, Qwen1.5-MOE-sft-nemotron-code is fine-tuned on code and reasoning datasets, which dramatically alter the model's generation patterns and distribution of attention. Such changes can inadvertently disrupt or overwrite safety neuron activations learned during pretraining or alignment. This aligns with prior findings in dense models, where benign fine-tuning has been shown to degrade safety alignment~\cite{qi2023fine}. While that conclusion was drawn initially for dense LLMs, our results extend it to MoE LLMs, highlighting that their modular structure does not inherently protect against safety erosion during downstream fine-tuning.

\subsection{Attack on MoE Vision Language Models}
\ourname generalizes beyond purely text-based MoE LLMs and is effective against modern MoE-based Vision Language Models (VLMs) as well. In this section, we evaluate its applicability to two prominent 2025 VLM families: the Deepseek-VL2 series~\cite{wu2024deepseek} and the Kimi-VL family~\cite{team2025kimi}. Both adopt a mixture MoE architecture (Fig.~\ref{subfig:mixture_moe}) for the language model component, augmented with a vision encoder to handle image inputs. We first apply the full \ourname pipeline to Deepseek-VL2 and Kimi-VL-A3B-Instruct. Then, to evaluate cross-model transferability in the multi-modal setting, we perform a one-shot transfer attack: the safety neurons identified from Kimi-VL-A3B-Instruct are directly applied to attack two reasoning-enhanced variants: Kimi-VL-A3B-Thinking and Kimi-VL-A3B-Thinking-2506. For the dataset, we convert malicious prompts from the StrongREJECT dataset~\cite{souly2024strongreject} to images \nadd{with printed texts}, following previous works~\cite{gong2025figstep,weng2024textit}.\footnote{We have also performed experiments with 500 not-safe-for-work (NSFW) images, asking the model to describe the image. Results show that all VLMs are unaligned with NSFW content, as all images are described without a single refusal.}
\begin{table}[ht]
\centering
\footnotesize
\begin{tabular}{l|cc|c}
\toprule
\textbf{Target Model} &\textbf{0\%} & \textbf{100\%} & \textbf{Ratio}\\
\midrule
Deepseek-VL2-Small & 42.7\% & 64.6\% & 2.6\% \\
Deepseek-VL2 & 23.6\% & 64.0\% & 2.5\% \\
Kimi-VL-A3B-Instruct & 9.9\% & 64.3\% & 2.3\% \\
Kimi-VL-A3B-Thinking (T) & 16.8\% & 54.0\% & 2.3\%\\
Kimi-VL-A3B-Thinking-2506 (T) & 10.8\% & 57.8\% & 2.3\% \\
\midrule
\emph{Average} & \emph{20.8\%} & \emph{60.9\%} & \emph{2.4\%}  \\
\bottomrule
\end{tabular}
\caption{ASR with MoE VLM.}
\label{tab:attack vlm moe}
\end{table}

The attack results are shown in Table~\ref{tab:attack vlm moe}; ``T'' denotes the one-shot transfer attack. The results demonstrate that \ourname remains highly effective in this multi-modal setting. Across all models, \ourname significantly improves ASR from 20.8\% to 60.9\%. Notably, even the one-shot transfer attack achieves strong performance, pushing ASR above 54\% on both Kimi-VL reasoning variants. These results suggest that safety alignment in VLMs is still mediated largely through the MoE language component, and that the vulnerabilities \ourname targets persist in multi-modal extensions. The average safety neuron ratio remains low at 2.4\%, consistent with our findings in unimodal models.

\subsection{Performance Benchmark}
We benchmark \ourname against SAFEx~\cite{lai2025safex}, the only prior work, to our knowledge, that proposes an inference-time attack on MoE LLMs. While SAFEx originally evaluates only four models, we reproduce their expert-pruning approach across all eight models listed in Table~\ref{tab:moe specs} for a fair and comprehensive comparison. Following SAFEx's methodology, we use 20\,000 harmful requests (typically refused by aligned models) and 20\,000 jailbreak prompts (crafted to bypass refusal) to profile and select the target experts for pruning~\cite{luo2024jailbreakv28k}.

\begin{table}[ht]
\centering
\footnotesize
\begin{tabular}{l|cc}
\toprule
\textbf{Target Model} & \textbf{SAFEx}~\cite{lai2025safex} & \textbf{\ourname} \\
\midrule
GPT-OSS-20B & 6.4\% & 80.2\% \\
Qwen3-30B-A3B-Instruct-2507 & 28.4\% & 56.9\% \\
Phi-3.5-MoE-Instruct & 26.8\% & 56.5\% \\
Mixtral-8x7B-Instruct-v0.1 & 48.8\% & 64.2\% \\
Qwen1.5-MoE-A2.7B-Chat & 35.0\% & 57.8\% \\
Deepseek-MoE-16b-Chat & 35.6\% & 53.6\% \\
Hunyuan-A13B-Instruct & 28.4\% & 76.9\% \\
OpenPangu-Pro-MoE-72B & 30.0\% & 73.1\% \\
\midrule
\emph{Average} & \emph{29.9\%} & \emph{64.9\%}\\
\bottomrule
\end{tabular}
\caption{ASR Benchmark with Existing Method.}
\label{tab:performance benchmark}
\end{table}

As shown in Table~\ref{tab:performance benchmark}, \ourname substantially outperforms SAFEx across all evaluated models, achieving an average ASR of 64.9\%, more than double the 29.9\% of SAFEx. The performance gap is particularly striking for models like GPT-OSS-20B, where \ourname increases ASR from just 6.4\% to 80.2\%. Even in models where SAFEx performs moderately well, such as OpenPangu-Pro-MoE-72B (30.0\%), \ourname still reaches a significantly higher ASR of 73.1\%. While SAFEx achieves a high ASR on Mixtral-8x7B (48.8\%), \ourname improves upon it with 64.2\%, indicating consistently stronger attack effectiveness. These improvements highlight several key advantages of our method: (i) \ourname operates at the neuron level, allowing fine-grained targeting of safety-critical components within experts; (ii) our profiling distinguishes between harmful and benign prompts using detailed activation patterns rather than binary jailbreak labels; and (iii) \ourname modifies only a small subset of neurons within the model, avoiding the coarse and destructive removal of entire experts. These advantages make \ourname more precise and effective than its counterpart.

\subsection{Utility Analysis}
To verify whether \ourname compromises the broader utility of the model, we evaluate both unmodified and pruned versions on five established natural language understanding (NLU) and reasoning benchmarks listed in Section~\ref{subsec:Evaluation Metrics}.
\begin{table*}[ht]
\centering
\footnotesize
\begin{tabular}{l|cc|cc|cc|cc|cc}
\toprule
\textbf{Target Model} & 
\multicolumn{2}{c|}{\textbf{CoLA}} & 
\multicolumn{2}{c|}{\textbf{RTE}} & 
\multicolumn{2}{c|}{\textbf{WinoGrande}} & 
\multicolumn{2}{c|}{\textbf{OpenBookQA}} & 
\multicolumn{2}{c}{\textbf{ARC}} \\
 & \textbf{Before} & \textbf{After} 
 & \textbf{Before} & \textbf{After} 
 & \textbf{Before} & \textbf{After} 
 & \textbf{Before} & \textbf{After} 
 & \textbf{Before} & \textbf{After} \\
\midrule
GPT-OSS-20B & 69.7\% & 69.7\% & 83.4\% & 84.6\% & 69.4\% & 62.2\% & 32.4\% & 29.2\% & 39.1\% & 38.5\% \\
Qwen3-30B-A3B-Instruct-2507 & 86.0\% & 86.2\% & 88.1\% & 87.4\% & 75.0\% & 75.5\% & 88.6\% & 74.2\% & 92.3\% & 92.3\% \\
Phi-3.5-MoE-Instruct & 84.0\% & 84.0\% & 85.4\% & 87.4\% & 13.0\% & 13.0\% & 35.2\% & 34.2\% & 63.9\% & 43.1\% \\
Mixtral-8x7B-Instruct-v0.1 & 84.6\% & 84.8\% & 85.9\% & 81.7\% & 64.6\% & 64.6\% & 72.0\% & 77.2\% & 79.3\% & 67.4\% \\
Qwen1.5-MoE-A2.7B-Chat & 79.9\% & 73.3\% & 75.8\% & 73.7\% & 55.5\% & 53.0\% & 73.8\% & 60.0\% & 72.6\% & 64.2\% \\
Deepseek-MoE-16b-Chat & 75.1\% & 51.3\% & 76.9\% & 75.5\% & 44.2\% & 42.0\% & 82.0\% & 76.4\% & 67.9\% & 70.1\%  \\
Hunyuan-A13B-Instruct & 36.2\% & 36.2\% & 60.7\% & 59.2\% & 31.3\% & 31.3\% & 34.4\% & 35.4\% & 33.1\% & 25.4\% \\
OpenPangu-Pro-MoE-72B & 84.0\% & 75.8\% & 89.9\% & 73.6\% & 49.6\% & 41.5\% & 30.0\% & 25.0\% & 27.1 & 21.1\% \\
\midrule
\emph{Average} & \emph{74.9\%} & \emph{70.2\%} & \emph{80.8\%} & \emph{77.9\%} & \emph{50.3\%} & \emph{47.9\%} & \emph{56.0\%} & \emph{51.4\%} & \emph{59.4\%} & \emph{52.8\%} \\
\bottomrule
\end{tabular}
\caption{Utility Evaluation on Five NLU Benchmarks Before and After Applying \ourname.}
\label{tab:utility_evaluation}
\end{table*}

As shown in Table~\ref{tab:utility_evaluation}, there is no significant drop in task performance after \ourname attack. The average accuracy on RTE decreased modestly from 80.8\% to 77.9\%, and CoLA remained stable with a 4.7 percentage point drop. Similar small reductions were observed for WinoGrande and OpenBookQA. ARC showed a slightly larger reduction (59.4\% to 52.8\%), driven mainly by performance drops in smaller or more sensitive models.
Interestingly, \ourname did not always harm performance; some models maintained or even improved their results on certain tasks. GPT-OSS-20B preserved its CoLA score and improved on RTE (83.4\% to 84.6\%). Qwen3-30B-A3B-Instruct-2507 gained slightly on CoLA and WinoGrande, while Mixtral-8x7B-Instruct showed consistent post-pruning gains on CoLA and OpenBookQA. 
These results suggest that \ourname does not broadly harm model utility. Instead, its effects depend on the model architecture; some models even benefit slightly from the removal of safety constraints. Overall, \ourname preserves general language and reasoning capabilities across diverse MoE models. Its targeted intervention disables safety alignment without significantly affecting standard task performance, and in some cases, may even reveal underlying capabilities that were previously suppressed.

\section{Ablation and Hyperparameter Study}
\label{sec:ablation study}
\nadd{We investigate the impact of experts, linear layers, safety expert range, and $z$-threshold in this section. The correlation between the identified neurons and the model's refusal behavior is studied in Appendix~\ref{apdsec:pruning strength} and~\ref{apdsec:safety neurons}.}

\subsection{Sparse vs. Shared Experts}
\label{subsec:Sparse vs. Shared Experts}
Recall that two types of experts, sparse and shared experts, exist in mixture and grouped mixture MoEs. 
To understand how different types of experts contribute to safety alignment in MoE architectures, we perform an ablation study that isolates the safety impact of sparse and shared experts. We apply targeted safety removal separately to each group and measure the resulting ASR. 
\begin{table}[ht]
\centering
\footnotesize
\begin{tabular}{l|ccc}
\toprule
\textbf{Target Model} & \textbf{Sparse} & \textbf{Shared} & \textbf{All} \\
\midrule
Qwen1.5-MoE-A2.7B-Chat & 13.1\% & 42.2\% & 57.8\% \\
Deepseek-MoE-16b-Chat & 48.6\% & 14.1\% & 53.6\% \\
Hunyuan-A13B-Instruct & 68.4\% & 39.9\% & 76.9\% \\
OpenPangu-Pro-MoE-72B & 24.9\% & 50.2\% & 73.1\% \\
\midrule
\emph{Average} & \emph{38.8\%} & \emph{36.6\%} & \emph{65.3\%} \\
\bottomrule
\end{tabular}
\caption{Safety Contribution of Different Types of Experts.}
\label{tab:expert-type-study}
\end{table}

As shown in Table~\ref{tab:expert-type-study}, pruning sparse experts yields a slightly higher average ASR increase (38.8\%) compared to pruning shared experts (36.6\%). However, this observation masks an important asymmetry: shared experts are few in number but always active for all tokens, meaning their influence on model behavior is disproportionately high relative to their parameter count. In contrast, sparse experts are many, but each is only activated for a subset of tokens, depending on routing. Therefore, when considering the \emph{per-expert impact on safety}, shared experts are critical for the model's safety. 

This distinction is most evident in OpenPangu-Pro-MoE-72B, where pruning just the shared expert results in a 50.2\% ASR, nearly matching the 57.8\% ASR observed when pruning both expert types. Similarly, in Qwen1.5-MoE-A2.7B-Chat, shared expert pruning alone yields a 42.2\% ASR, triple the impact compared with sparse expert pruning (13.1\%). These results suggest that although sparse experts carry distributed safety functionality across many layers, shared experts often serve as centralized, always-on enforcement points. Conversely, in models like Deepseek-MoE-16b-Chat and Hunyuan-A13B-Instruct, sparse experts play a more dominant role, far exceeding the corresponding shared expert contributions. This variation suggests that the reliance on shared versus sparse experts for implementing safety may be model-dependent, possibly influenced by the design of the MoE routing mechanism or the alignment strategy used during training. 

\subsection{Gate-Projection vs. Up-Projection Layers}
\label{subsec:Gate vs. Up-Projection Layers}
To further localize where safety behavior resides within an expert, we conduct an ablation study that isolates the contributions of two key sublayers in each expert’s MLP block: the gate-projection and the up-projection layer. The down-projection layer is excluded as it maps back to the model's hidden dimension and generally compresses information rather than introducing specialized behavior~\cite{nelson2021mathematical}.
\begin{table}[ht]
\centering
\footnotesize
\begin{tabular}{l|ccc}
\toprule
\textbf{Target Model} & \textbf{Gate-Proj.} & \textbf{Up-Proj.} & \textbf{All} \\
\midrule
GPT-OSS-20B & 68.7\% & 10.5\% & 80.2\% \\
Qwen3-30B-A3B-Instruct-2507 & 50.5\% & 1.6\% & 56.9\% \\
Phi-3.5-MoE-Instruct & 48.9\% & 2.8\% & 56.5\% \\
Mixtral-8x7B-Instruct-v0.1 & 59.1\% & 30.7\% & 64.2\% \\
Qwen1.5-MoE-A2.7B-Chat & 49.2\% & 16.6\% & 57.8\% \\
Deepseek-MoE-16b-Chat & 29.7\% & 27.8\% & 53.6\% \\
Hunyuan-A13B-Instruct & 70.3\% & 50.2\% & 76.9\% \\
OpenPangu-Pro-MoE-72B & 70.4\% & 24.3\% & 73.1\% \\
\midrule
\emph{Average} & \emph{55.9\%} & \emph{20.6\%} & \emph{64.9\%} \\
\bottomrule
\end{tabular}
\caption{Safety Contribution of Different Types of Layers.}
\label{tab:prune-layer-study}
\end{table}

Table~\ref{tab:prune-layer-study} presents the ASR after pruning safety neurons from (i) the gate-projection layer only, (ii) the up-projection layer only, and (iii) both layers combined. On average, pruning neurons in the gate layer alone increases ASR to 55.9\%, while pruning the up-projection layer alone increases ASR only to 20.6\%. When both are pruned, ASR reaches 64.9\%, confirming that both layers contribute, but with dramatically different magnitudes.
The dominance of the gate-projection layer in safety behavior is both empirically evident and architecturally intuitive. The gate layer applies the first learned transformation to the input token representation. This early-stage computation often determines whether certain high-level semantic features are activated or suppressed. Safety behaviors, such as recognizing harmful intent or initiating refusal patterns, likely depend on these early activations being triggered correctly. Besides, neurons in the gate layer typically exhibit sparser and more discriminative activations, especially in transformer MLPs~\cite{geva2022transformer}. This sparsity makes them ideal candidates for implementing specialized behaviors, such as safety responses, which activate only when harmful intent is detected.

\subsection{The Range of Safety Experts}
\label{subsec:The Range of Safety Experts}
The safety experts guide the neuron-level analysis and pruning procedure. However, the number of safety experts to include remains an important hyperparameter: selecting too few may miss key safety pathways, while including too many could dilute the precision of the attack. To study this trade-off, we conduct a hyperparameter study where we vary the number of safety experts used for pruning. 
\begin{table}[ht]
\centering
\footnotesize
\begin{tabular}{l|ccc}
\toprule
\textbf{Target Model} & \textbf{Top-$k$} & \textbf{Top-$2k$} & \textbf{Top-$3k$} \\
\midrule
GPT-OSS-20B & 49.8\% & 78.9\% & 80.2\% \\
Qwen3-30B-A3B-Instruct-2507 & 26.5\% & 47.3\% & 56.9\% \\
Phi-3.5-MoE-Instruct & 5.8\% & 39.2\% & 56.5\% \\
Mixtral-8x7B-Instruct-v0.1 & 44.7\% & 45.4\% & 64.2\% \\
Qwen1.5-MoE-A2.7B-Chat & 48.2\% & 57.2\% & 57.8\% \\
Deepseek-MoE-16b-Chat & 30.0\% & 40.9\% & 53.6\% \\
Hunyuan-A13B-Instruct & 28.4\% & 33.2\% & 76.9\% \\
OpenPangu-Pro-MoE-72B & 62.9\% & 68.0\% & 73.1\% \\
\midrule
\emph{Average} & \emph{37.0\%} & \emph{51.3\%} & \emph{64.9\%} \\
\bottomrule
\end{tabular}
\caption{ASR with different numbers of safety experts.}
\label{tab:safety-expert-study}
\end{table}

As shown in Table~\ref{tab:safety-expert-study}, increasing the number of safety experts consistently improves ASR. On average, ASR improves from 37.0\% (Top-$k$) to 51.3\% (Top-$2k$), and reaches 64.9\% with Top-$3k$. Aligned with the observation in Section~\ref{sec:case study}, this suggests that safety behavior is not tightly localized to just a few (e.g., the most frequently activated) experts per token. Instead, it is distributed across a broader subset of experts. 
Notably, the marginal gain between top-$2k$ and top-$3k$ is still meaningful (e.g., 68.0\% to 73.1\% in OpenPangu, 47.3\% to 56.9\% in Qwen3), indicating that even among relatively less frequently selected experts, some contribute critically to safety enforcement. Selecting a wider range of experts enables \ourname to more fully capture the safety-related subspace of the model: 2.6\% of neurons per layer on average.

\subsection{Impact of the $z$-Threshold}
\label{subsec:Impact of the Z-Threshold}

\ourname identifies safety-relevant neurons by selecting outliers in their safety weights, quantified via $z$-scores that capture how strongly a neuron's activation distinguishes between harmful and benign prompts. In this section, we investigate the influence of the $z$-score threshold $\tau$ on the attack.
\begin{table}[ht]
\centering
\footnotesize
\begin{tabular}{l|ccc}
\toprule
\textbf{Target Model} & \textbf{$\tau=1$} & \textbf{$\tau=2$} & \textbf{$\tau=3$} \\
\midrule
GPT-OSS-20B & 79.2\% & 80.2\% & 57.8\% \\
Qwen3-30B-A3B-Instruct-2507 & 73.8\% & 56.9\% & 28.8\% \\
Phi-3.5-MoE-Instruct & 72.8\% & 56.5\% & 38.0\% \\
Mixtral-8x7B-Instruct-v0.1 & 64.9\% & 64.2\% & 52.7\% \\
Qwen1.5-MoE-A2.7B-Chat & 67.7\%* & 57.8\% & 38.3\% \\
Deepseek-MoE-16b-Chat & 53.4\% & 53.6\% & 42.5\% \\
Hunyuan-A13B-Instruct & 72.5\% & 76.9\% & 30.0\% \\
OpenPangu-Pro-MoE-72B & 88.5\%* & 73.1\% & 52.1\% \\
\midrule
\emph{Average} & \emph{71.6\%} & \emph{64.9\%} & \emph{42.5\%} \\
\bottomrule
\end{tabular}
\caption{ASR with different $z$ Threshold.}
\label{tab:zscore-study}
\end{table}

Table~\ref{tab:zscore-study} presents the ASR across eight MoE LLMs using three representative thresholds: $\tau = 1$, $\tau = 2$ (our default), and $\tau = 3$. The ASR with a ``*'' indicates some responses are collapsed, e.g., repeating certain words or sentences.
Lower thresholds ($\tau=1$) generally achieve higher ASR by suppressing more neurons, but risk removing non-safety neurons and destabilizing generation (e.g., Qwen1.5-MoE and OpenPangu). Higher thresholds ($\tau=3$) significantly reduce ASR across all models, suggesting insufficient suppression of safety mechanisms. The default threshold ($\tau=2$) offers the best balance, yielding consistently strong ASR while preserving model utility. Interestingly, models such as DeepSeek-MoE and Mixtral show stable ASR across thresholds, indicating distributed safety representations, whereas Qwen3-30B and GPT-OSS-20B are more sensitive to \ourname attack, suggesting more localized safety neurons.

\section{Discussion}
\label{sec:discussion}

\textbf{Black-box Attack.}
While \ourname is designed for white-box settings, its insights can guide black-box attacks via proxy models. One practical path is to use a publicly available MoE model (e.g., GPT-OSS-20B) to identify expert activation patterns associated with safety alignment. Once safety-relevant experts are identified, an attacker can craft adversarial prompts that deliberately avoid triggering those experts (e.g., by avoiding tokens or phrasing that typically activate them). These prompts can then be deployed against closed-source or black-box MoE models assuming shared structural or alignment similarities, an assumption supported by our transfer attacks across sibling models (see Section~\ref{subsec:transfer attack}). This strategy enables targeted prompt manipulation without needing access to the internal states of the black-box model, much like jailbreak attacks optimized through reinforcement learning or optimization in proxy environments~\cite{zou2023universal,wu2025neurostrike}.

\textbf{Potential Defenses.}
Most existing defense research has focused on dense LLMs. Techniques such as reinforced safety tuning~\cite{ouyang2022training} and adversarial training~\cite{zou2023universal} are designed and evaluated primarily on dense models. As a result, there is currently no widely adopted or evaluated defense specifically tailored to MoE LLMs. 
Our findings highlight a structural vulnerability unique in current MoE-based LLMs: safety alignment tends to concentrate in a small subset of neurons, which can be surgically disabled.
To harden the MoE LLMs, one approach is to enforce \emph{safety redundancy} by explicitly distributing safety-aligned behavior across multiple experts during training. This could be achieved through load-balancing-aware alignment objectives or by encouraging diverse experts to respond similarly to harmful prompts (e.g., via contrastive alignment or regularization across experts). Next, instead of treating safety as an auxiliary property, future training objectives could incorporate safety constraints as a core component of every expert’s behavior. This may involve integrating refusal behavior more deeply into the reward modeling or fine-tuning stages (e.g., using RLHF or DPO~\cite{ouyang2022training, rafailov2023direct} applied to individual experts or expert ensembles). Finally, as \ourname relies on inference-time activation modification, service providers could monitor or checksum internal activations or critical layers. 

\section{Related Works}
\label{sec:related}

Research on LLM safety has largely focused on jailbreak attacks and defenses for dense architectures. Early jailbreaks relied on fixed prompt templates, including role-play, hidden directives, and obfuscation~\cite{perez2022ignore,jiang2023prompt,li2023deepinception,shen2024anything}, but these became less effective as models improved. More adaptive methods emerged, using fuzzing~\cite{yu2023gptfuzzer}, gradient-based prompt optimization~\cite{ebrahimi2017hotflip,wen2023hard}, or generative LLMs to iteratively refine adversarial inputs~\cite{zhou2022large,chao2023jailbreaking,chang2024play}. While these approaches increase flexibility, they remain input-level attacks and overlook the internal mechanisms underlying safety alignment. In parallel, neuron interpretability studies~\cite{kadar2017representation,lakretz2019emergence,anthropic-neurons} have shown that individual neurons can encode specific concepts or \nadd{safety-relevant features}, motivating neuron-targeted attack and defense strategies. \nadd{Some recent works show that removing safety-related neurons could dramatically increase the ASR~\cite{wei2024assessing,wu2025neurostrike}.}

On the other hand, safety analysis for MoE architectures is still limited despite their growing adoption in LLMs. Compared to dense models, MoEs distribute computation across multiple experts, which may introduce unique vulnerabilities that recent work has started to uncover. For example, Hayes et al.~\cite{hayes2024buffer} show that cross-batch routing strategies can be exploited for integrity and availability attacks, Wang et al.~\cite{wang2025badmoe} demonstrate backdoors by poisoning dormant experts, and Yona et al.~\cite{yona2024stealing} reveal a side-channel leakage attack that extracts user prompts via routing tie-breaks. Among the existing works, Lai et al. propose SAFEx~\cite{lai2025safex} that identifies and then masks safety control experts to reduce the refusal rates. However, it operates at the coarse expert level and assumes the existence of dedicated safety experts. In contrast, \ourname performs a fine-grained neuron-level attack, enabling precise and effective safety removal with limited modification.
\section{Conclusions}
\label{sec:conclusions}

We present \ourname, the first inference-time attack framework that systematically compromises safety alignment in MoE LLMs with mainstream architectures: sparse, mixture, and grouped mixture.
\ourname introduces a lightweight three-stage attack pipeline: (i) gate-level profiling to identify safety-critical experts disproportionately activated by harmful prompts, (ii) expert-level localization to isolate safety neurons whose removal suppresses refusal behavior, and (iii) target safety removal to compromise the safety alignment. Across eight open MoE models, \ourname increases the average ASR from 7.4\% to 64.9\% by removing 2.6\% of neurons in the relevant expert layers on average. \ourname generalizes to five MoE VLMs, increasing ASR from 20.8\% to 60.9\%. Moreover, these neurons transfer robustly across variants within the same family, achieving 67.7\% ASR in one-shot transfer attacks. As MoE LLMs become foundational infrastructure for AI deployment, our work highlights a critical and previously overlooked security gap in MoE models. 

\section*{Acknowledgement}
Our research work was partially funded by DFG-SFB 1119-236615297, the European Union under Horizon Europe Programme-Grant Agreement 101070537-CrossCon and-Grant Agreement 101093126-ACES, NSF-DFG-Grant 538883423, the European Research Council under the ERC Programme-Grant 101055025-HYDRANOS, as well as the Federal Ministry of Education and Research of Germany (BMBF) within the IoTGuard project. Any opinions, findings, conclusions, or recommendations expressed herein are those of the authors and do not necessarily reflect those of the European Union, the European Research Council, or the Federal Ministry of Education and Research of Germany.
\label{sec:ccknowledgement}
\newpage
\section*{Ethical Considerations}

This work investigates the structural safety mechanisms of Mixture-of-Experts (MoE) language models and introduces \ourname, a framework for identifying and selectively removing safety-related computations at inference time. We carefully considered the ethical implications of this research, particularly its dual-use nature. While our findings could potentially be misused to weaken or bypass existing safety alignment mechanisms, they also provide critical insights for model developers, security researchers, and the broader AI safety community into a class of vulnerabilities that remains largely underexplored.

\noindent\textbf{Stakeholders.}
The primary stakeholders affected by this research are organizations that develop and deploy large language models, particularly those adopting Mixture-of-Experts (MoE) architectures. Our evaluation focuses on openly released MoE models from a range of major developers, including OpenAI, Alibaba, Microsoft, Mixtral, DeepSeek, Tencent, and Huawei. We have notified relevant stakeholders of our framework and our findings and are following coordinated vulnerability disclosure practices. Where appropriate, we will incorporate feedback from affected parties and adhere to any reasonable disclosure timelines. This process is intended to ensure that the insights presented in this work contribute constructively to improving the safety and robustness of MoE-based systems.

\noindent\textbf{Potential Impact.}
We acknowledge that \ourname could, in principle, be misused by malicious actors to weaken or bypass safety alignment mechanisms. Such risks underscore the importance of responsible use and contextual understanding of the framework. To mitigate potential misuse, we clearly communicate the defensive and diagnostic purpose of our work. The ultimate goal is to strengthen security and inform the development of future safety mechanisms, ensuring that MoE-based systems are more resilient before deployment.

\noindent\textbf{Mitigations for Negative Impacts.}
This work does not involve human subjects, personal data, or interaction with deployed systems. All experiments were conducted on openly released models in controlled research environments.

To minimize the risk of misuse, we release only the components necessary for reproducibility and scientific evaluation. The accompanying artifact will include a clear “research-only” disclaimer and usage guidelines emphasizing that the framework is intended solely for defensive and diagnostic purposes. In Section~\ref{sec:discussion} of the paper, we discussed the potential defenses that can effectively mitigate the \ourname framework. By promoting responsible disclosure and controlled dissemination, we aim to ensure that our work advances safety research without enabling harmful applications.

\noindent\textbf{Decision to Conduct and Publish the Study.}
We chose to conduct and publish this research because safety defenses for MoE language models lag significantly behind those for dense architectures, despite the growing adoption of MoE designs. Leaving these vulnerabilities unexamined could pose substantial risks as such models see broader deployment. We believe that the benefits of transparency, early disclosure, and catalyzing defensive research outweigh the potential harms of publication, particularly when combined with responsible disclosure and controlled dissemination.

\noindent\textbf{Protection of Research Team Members.}
The research team carefully considered the psychological and ethical challenges involved in examining model outputs that may contain sensitive or harmful content. While we involved a human judge for a small part of our research to rule out false positives in harmfulness detection, all such evaluations were performed in controlled settings with clear safety protocols. Team members were briefed on the potential exposure to disturbing or offensive content and were encouraged to discontinue participation at any time if they experienced discomfort. We also established procedures for reporting distress and provided access to mental health and well-being resources. These measures ensured that the research was conducted responsibly while safeguarding the well-being of all contributors.

\section*{Open Science}
In alignment with USENIX Security’s open science policy, we make all artifacts from this study permanently available to the security research community (\url{https://doi.org/10.5281/zenodo.17910455}). This includes the source code for gate-level profiling, expert-level localization, and targeted safety removal. \ourname exclusively targets openly released models with public datasets, enabling full replication of our methodology and results. The entire pipeline is designed to run on consumer-grade GPU, ensuring that independent researchers can validate and extend our work without access to large-scale computing infrastructure.

\bibliographystyle{IEEEtran}
\bibliography{bibliography}

\newpage
\appendix

\section*{Appendix} 
\label{apd:appendix}

\section{Sparse Expert Ablation}
\label{apdsec:Expert Ablation}

\nadd{As stated in Section~\ref{subsec:moe-archi}, the router sparsely activates the top-K experts in each MoE layer. Since this selection strongly influences the model’s response to different prompts, we treat K as a key hyperparameter.} We conduct sparse expert ablation by progressively removing the top-0.5K, top-K, and top-2K experts per layer, ranked by their malicious utility scores. As shown in Table~\ref{tab:attacks moe experts}, this targeted ablation consistently increases the Attack Success Rate (ASR) across all evaluated MoE LLMs, demonstrating the safety relevance of the most active experts during harmful input processing. Overall, the average ASR increases from 19.3\% (top-0.5K) to 45.9\% (top-2K), confirming that experts with high malicious utility scores play a central role in mediating refusal behavior. 
\begin{table}[ht]
\centering
\footnotesize
\begin{tabular}{l|ccc}
\toprule
\textbf{Target Model} &\textbf{Top-0.5K} & \textbf{Top-K} & \textbf{Top-2K}\\
\midrule
GPT-OSS-20B & 15.0\% & 32.9\% & 73.8\% \\
Qwen3-30B-A3B-Instruct-2507 & 4.8\% & 26.5\% & 39.6\% \\
Phi-3.5-MoE-Instruct & 0.0\% & 47.3\% & 58.8\% \\
Mixtral-8x7B-Instruct-v0.1 & 56.9\% & 21.7\% & 33.5\% \\
Qwen1.5-MoE-A2.7B-Chat & 10.9\% & 10.5\% & 15.7\% \\
Deepseek-MoE-16b-Chat & 47.9\% & 70.0\% & 57.2\% \\
Hunyuan-A13B-Instruct & 11.8\% & 14.1\% & 71.9\% \\
OpenPangu-Pro-MoE-72B & 6.7\% & 8.6\% & 16.6\% \\
\midrule
\emph{Average} & \emph{19.3\%} & \emph{29.0\%} & \emph{45.9\%} \\
\bottomrule
\end{tabular}
\caption{ASR with Sparse Expert Ablation.}
\label{tab:attacks moe experts}
\end{table}

Notably, the ASR gains are not uniform across models. For example, GPT-OSS-20B shows a dramatic rise from 15.0\% (top-0.5K) to 73.8\% (top-2K), highlighting a strong concentration of safety behavior in a relatively small subset of experts. Similarly, Hunyuan-A13B-Instruct and Phi-3.5-MoE-Instruct also show steep increases at the top-2K level, reaching 71.9\% and 58.8\%, respectively. These results indicate that safety alignment in MoE models is not widely distributed, but rather localized to specific, highly utilized experts.

Interestingly, Mixtral-8x7B-Instruct-v0.1 exhibits an unusually high ASR even at the top-0.5K level (56.9\%), but a more modest increase across the broader ablation range, suggesting that a small number of experts may disproportionately influence safety outcomes in that model. In contrast, Qwen1.5-MoE-A2.7B-Chat and OpenPangu-Pro-MoE-72B show more gradual increases, which may reflect a more diffused or redundant safety structure.

\nadd{\section{Safety Neuron Suppression Strength}
\label{apdsec:pruning strength}
To understand how strongly safety neurons contribute to refusal behavior, we vary the suppression strength applied to identified safety neurons and measure the resulting ASR. Specifically, we evaluate partial activation suppression at 35\% and 65\%, and compare them against full suppression (100\%).
\begin{table}[ht]
\centering
\footnotesize
\begin{tabular}{l|ccc}
\toprule
\textbf{Target Model} & \textbf{35\%} & \textbf{65\%} & \textbf{100\%} \\
\midrule
GPT-OSS-20B & 9.9\% & 68.6\% & 80.2\% \\
Qwen3-30B-A3B-Instruct-2507 & 2.2\% & 22.6\% & 56.9\% \\
Phi-3.5-MoE-Instruct & 1.5\% & 11.5\% & 56.5\% \\
Mixtral-8x7B-Instruct-v0.1 & 36.7\% & 54.6\% & 64.2\% \\
Qwen1.5-MoE-A2.7B-Chat & 25.6\% & 42.5\% & 57.8\% \\
Deepseek-MoE-16b-Chat & 25.2\% & 29.4\% & 53.6\% \\
Hunyuan-A13B-Instruct & 42.2\% & 66.1\% & 76.9\% \\
OpenPangu-Pro-MoE-72B & 35.8\% & 64.9\% & 73.1\% \\
\midrule
\emph{Average} & \emph{22.4\%} & \emph{45.0\%} & \emph{64.9\%} \\
\bottomrule
\end{tabular}
\caption{ASR with different numbers of safety experts.}
\label{tab:safety neuron suppression}
\end{table}

As shown in Table~\ref{tab:safety neuron suppression}, ASR increases monotonically with stronger suppression. On average, partial suppression at 35\% and 65\% yields ASRs of 22.4\% and 45.0\%, respectively, both substantially higher than the baseline ASR of 7.4\%, yet still noticeably lower than the 64.9\% achieved with full suppression. This trend is consistent across all evaluated MoE models, despite their architectural and routing differences.

These results indicate that safety behavior emerges from the collective contribution of multiple safety neurons. Partial suppression weakens refusal behavior proportionally, while full suppression is required to reliably disable it. The smooth degradation in safety performance further suggests a strong causal relationship between the activation magnitude of safety neurons and the model’s refusal behavior, rather than an all-or-nothing effect.}

\nadd{\section{The Importance of Safety Neurons}
\label{apdsec:safety neurons}
To verify that the observed effects are specific to safety neurons rather than a byproduct of indiscriminate neuron disruption, we compare targeted safety neuron suppression against random neuron pruning. We conduct two random baselines: \texttt{R1}, which prunes an equal number of neurons randomly selected from all experts, and \texttt{R2}, which randomly prunes neurons within the safety expert subsets.
\begin{table}[ht]
\centering
\footnotesize
\begin{tabular}{l|ccc}
\toprule
\textbf{Target Model} & \textbf{R1} & \textbf{R2} & \textbf{\ourname} \\
\midrule
GPT-OSS-20B & 1.3\% & 1.6\% & 80.2\% \\
Qwen3-30B-A3B-Instruct-2507 & 0.6\% & 0.6\% & 56.9\% \\
Phi-3.5-MoE-Instruct & 0.3\% & 0.6\% & 56.5\% \\
Mixtral-8x7B-Instruct-v0.1 & 12.8\% & 16.3\% & 64.2\% \\
Qwen1.5-MoE-A2.7B-Chat & 5.1\% & 7.3\% & 57.8\% \\
Deepseek-MoE-16b-Chat & 23.3\% & 18.5\% & 53.6\% \\
Hunyuan-A13B-Instruct & 12.5\% & 28.4\% & 76.9\% \\
OpenPangu-Pro-MoE-72B & 6.6\% & 21.1\% & 73.1\% \\
\midrule
\emph{Average} & \emph{7.8\%} & \emph{11.8\%} & \emph{64.9\%} \\
\bottomrule
\end{tabular}
\caption{ASR with random/safety neuron pruning.}
\label{tab:safety vs random neuron}
\end{table}
Table~\ref{tab:safety vs random neuron} shows that random neuron pruning results in consistently low ASR, with averages of 7.8\% and 11.8\%, closely matching the baseline behavior. In contrast, \ourname achieves a significantly higher average ASR of 64.9\%, with improvements exceeding 5$\times$ to 10$\times$ across most models. Indeed, the ineffectiveness of random pruning confirms that the observed jailbreak success is not due to reduced model capacity or degraded fluency, but rather to the targeted disruption of neurons that encode safety-relevant representations.}

\end{document}